\documentclass[onecolumn,prb,aps,notitlepage]{revtex4-1}
\usepackage{tikz}
\usepackage{amssymb}
\usepackage{psfrag}
\usepackage{textcomp}
\usepackage{pdfsync}
\usepackage{amsmath}
\usepackage{amsfonts}
\usepackage{graphicx,bm,epstopdf}
\usepackage{subfigure}
\usepackage{comment}
\usepackage{verbatim}
\usepackage{color}
\usepackage{upgreek}
\usepackage{tikz}
\usepackage{natbib,hyperref}
\usepackage{threeparttable}
\usepackage{ifpdf}
\usepackage{float}
\usepackage[normalem]{ulem}

\newcommand\redsout{\bgroup\markoverwith{\textcolor{red}{\rule[0.5ex]{2pt}{0.4pt}}}\ULon}
\usepackage{soul,xcolor}

\begin{document}
\setstcolor{red}

\title{Tunable mode coupling in nano-contact spin torque oscillators}

\author{Steven S.-L. Zhang$^{1,2}$}
\email{shulei.zhang@anl.gov}
\author{Ezio Iacocca$^{3,4}$}
\author{Olle Heinonen$^{2,5,6}$}
\email{heinonen@anl.gov}

\affiliation{$^1$Department of Physics and Astronomy, University of Missouri, Columbia, Missouri 65211, USA\\$^2$Material Science Division, Argonne National Laboratory, Lemont, Illinois 60439, USA\\$^3$Department of Applied Mathematics, University of Colorado, Boulder, Colorado 80309-0526, USA\\$^4$Department of Physics, Division for Theoretical Physics, Chalmers University of Technology, 412 96, Gothenburg, Sweden\\$^5$Northwestern-Argonne Institute of Science and Technology, 2145 Sheridan Road, Evanston, Illinois 60208, USA\\$^6$Computation Institute, The University of Chicago, 5735 S Ellis Ave, Chicago, Illinois 60637 USA}


\date{\today}

\begin{abstract}
Recent experiments on spin torque oscillators have revealed interactions between multiple magnetodynamic modes, including mode-coexistence, mode-hopping, and temperature-driven cross-over between modes. Initial multimode theory
has indicated that a linear coupling between several dominant modes, arising from the interaction of the subdynamic system with a magnon bath, plays an essential role in the generation of various multimode behaviors, such as mode hopping and mode coexistence.
In this work, we derive a set of rate equations to describe the dynamics of coupled magnetodynamic modes in a nano-contact spin torque oscillator. Expressions for both linear and nonlinear coupling terms are obtained, which allow us to analyze the dependence of the coupled dynamic behaviors of modes on external experimental conditions as well as intrinsic magnetic properties. For a minimal two-mode system, we further map the energy and phase difference of the two modes onto a two-dimensional phase space, and demonstrate in the phase portraits, how the manifolds of periodic orbits and fixed points vary with external magnetic field as well as with temperature.
\end{abstract}
\maketitle
\section{Introduction}

Since the discovery of the spin transfer torque (STT) effect~\cite{slonczewski1996jmmm,berger1996prb}, efficient manipulation of magnetization orientation can be achieved by applying a dc current perpendicularly to a magnetic heterostructure consisting of two magnetic layers separated by a nonmagnetic spacer. The current becomes spin polarized when passing through the magnetic layer with fixed magnetization direction and subsequently transfers spin angular momentum to the other magnetic layer by exerting a spin torque on the magnetization. One particularly important manifestation of the STT effect is the steady state magnetization dynamics at microwave frequencies that is realized in devices known as spin torque oscillators (STOs)~\cite{silva2008jmmm,Dumas2014}. These are typically fabricated in either nanopillar or nanocontact (NC) geometry and rely on the compensation of intrinsic damping by STT as the current approaches a threshold for auto-oscillations. With appropriate arrangement of the relative orientations of the magnetizations as well as of the current direction, nearly undamped oscillation modes with very small linewidth can be realized in STOs.

As a first attempt to describe the magnetodynamics in STOs, Slavin and coworkers~\cite{slavin2009ieee} developed a single-mode theory under the assumption that only a single coherent precessional mode is excited, which captures some remarkable nonlinear features of STOs qualitatively and to some extent quantitatively. The assumption of single-mode precession further precludes chaos and the possibility of mode transitions between dynamical modes~\cite{Bertotti2005}. Later, an effective theory of a two-mode STO was put forth by de Aguiar, Azevedo, and Rezende~\cite{deAguiar2007prb}. By solving the equations of motion for the amplitudes of the two modes, nonlinearly coupled by third order terms originating from four-magnon interactions, they concluded that, in steady state, only one mode will survive whereas the other will be extinguished. However, neither of these theories mentioned above can explain recent experimental observations of a variety of multi-mode dynamical effects in STOs such as mode-hopping~\cite{krivorotov2008prb,muduli2012prl,Muduli12PRB_T-dep-STO}, periodic mode transitions~\cite{bonetti2010prl,bonetti2012prb}, and mode coexistence~\cite{dumas2013prl,Iacocca2015prb}. In addition to mode-hopping, Muduli {\em et al.}\cite{Muduli12PRB_T-dep-STO} also noted a mode cross-over driven by {\em temperature} with other parameters, such as current and external magnetic field, kept fixed. Clearly, such temperature-driven behavior points to a highly non-trivial temperature dependence which must be explained by more comprehensive theories.

A multi-mode theory was first proposed by Muduli, Heinonen, and \AA{}kerman~\cite{muduli2012prl} to explain the observed mode hopping in nanopillar STOs. The authors showed, for a minimal two-mode system, that the rate equations for the slowly-varying mode amplitudes can be mapped onto a two-dimensional $Z_2-$symmetric dynamical system, in analogy with those for two counter-propagating modes in semiconductor ring lasers~\cite{beri2008prl,vanderSande}. A key ingredient of the theory is the assumption that there exists, in addition to third order nonlinear coupling terms~\cite{deAguiar2007prb}, a linear coupling between the modes, which is essential for the mode hopping to occur. By treating the various coupling coefficients in the rate equations as phenomenological parameters, the effective multimode theory has been well substantiated by later experimental observations that are related to mode hopping, including linewidth broadening in NC-STOs~\cite{Iacocca2014prb} and $1/f$-frequency noise spectrum in STOs~\cite{Sharma2014APL,Eklund2014}.

Effective control of STOs, however, requires in-depth understanding of the underlying physics of the mode coupling. For this purpose, the multimode theory was derived rigorously~\cite{Zhang16jmmm} from the micromagnetic Landau-Lifshitz-Gilbert equation, whereby the crucial linear coupling term was shown to arise from the interaction of a dynamical subsystem, which involves several dominant modes, with a thermal bath of magnons. This theoretical assertion is consistent with the two kinds of mode-coupling mechanisms that were identified experimentally in NC-STOs by Iacocca \emph{et al.}~~\cite{Iacocca2015prb}, namely magnon mediated scattering and intermode interaction. In this paper, we apply the multimode theory to a NC-STO for which approximate analytic profiles of the eigenmodes are available~\cite{slonczewski1999jmmm} and hence simplified expressions for both linear and nonlinear coupling terms can be derived explicitly. With these expressions, we further determine the dependence of the mode coupling and the ensuing dynamics of the STO on typical controllable experimental parameters such as the external magnetic field and temperature.

 The remainder of the paper is organized as follows. In Sec.~II, we outline the derivation of the coupled rate equations for the two lowest lying eigenmodes of a NC-STO, where the linear coupling term appears after the thermal bath of magnons is integrated out and the equations are projected onto the subspace of the two modes. We present explicitly the expressions for both linear and nonlinear coupling terms, and in particular show the dependence of the linear coupling term on the external magnetic field and temperature. We also reveal the correlation between the linear coupling and the nonlinear spin wave frequency shift. In Sec.~III, we transform the rate equations to a more appealing form, which allows mapping of the energy and phase difference of the two modes onto a two-dimensional phase space. We show in the resulting phase portraits how the manifolds of periodic orbits and fixed points, both stable and unstable ones, vary with external magnetic field as well as temperature. Finally, we discuss and summarize our results in Sec.~IV.
\section{Mode equations with linear coupling}

\begin{figure}[h]
\includegraphics[trim={0cm 0cm 0cm 0cm},clip=true, width=0.5\linewidth]{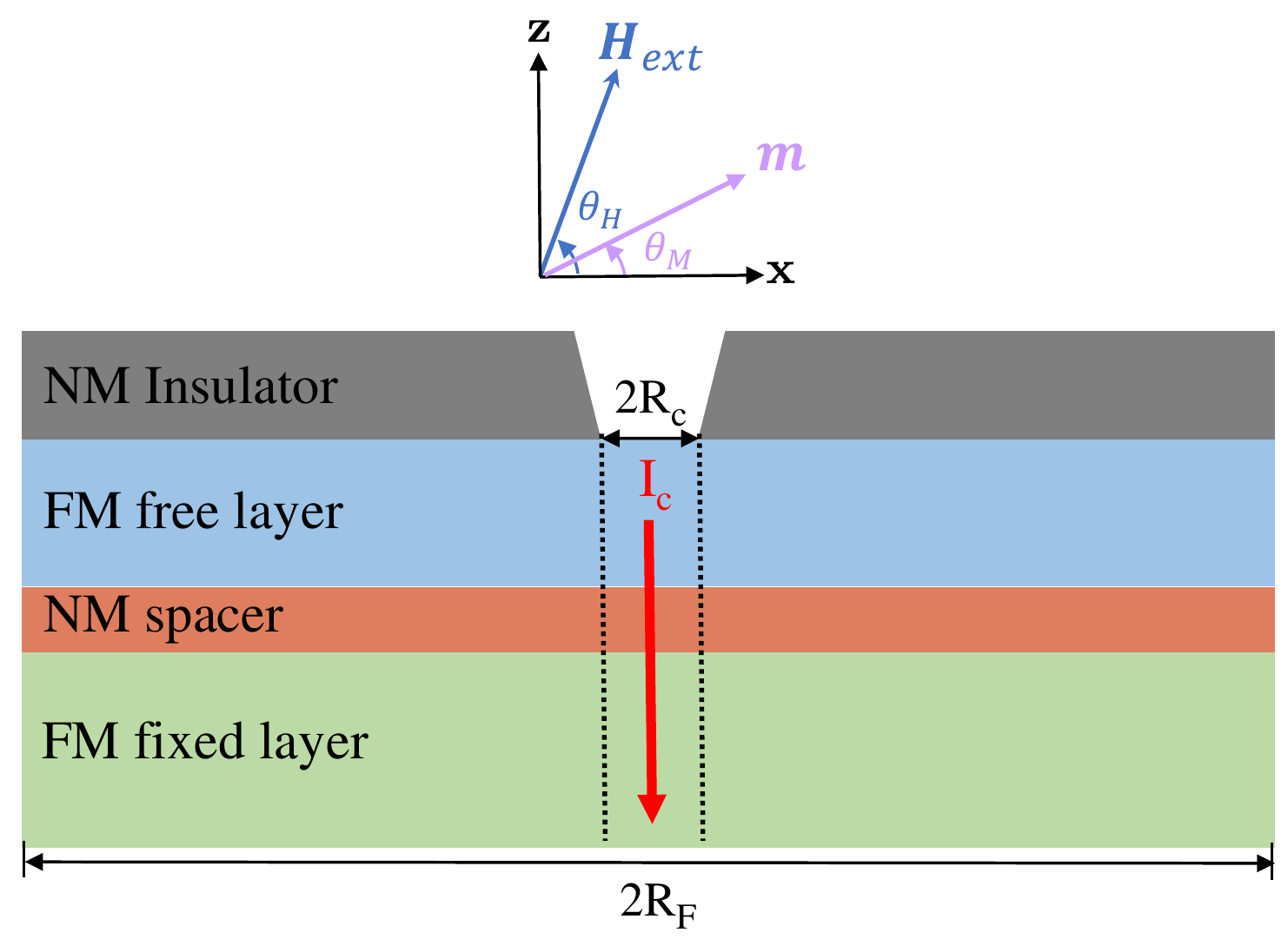}
\caption{Schematic diagram of the side view of a nano-contact spin torque oscillator. }
\label{fig:STO-schematics}
\end{figure}

We consider a NC-STO based on a pseudo spin valve composed of ferromagnetic (FM) fixed and free layers separated by a metallic nonmagnetic (NM) spacer, as shown in Fig.~\ref{fig:STO-schematics}. We further assume that the pseudo spin valve is patterned into a disc of radius $R_F$. A nanocontact of radius $R_c$ is defined on top of the free layer such that $R_c\ll R_F$. This configuration allows a current to flow through a cylindrical region directly below the nanocontact, which has been shown to be in good agreement with experiments~\cite{Dumas2014}.

The magnetization dynamics in a nanocontact spin torque oscillator can be
described by the generalized Landau-Lifshitz equation with current-induced spin transfer torque~\cite{slonczewski1996jmmm,berger1996prb},
i.e.,
\begin{equation}
\frac{\partial \mathbf{m}}{\partial t}=-\gamma \mathbf{m\times H}_\mathrm{eff}-%
\frac{\gamma \alpha }{1+\alpha ^{2}}\mathbf{m\times }\left( \mathbf{m\times H%
}_\mathrm{eff}\right) +\gamma a_{J}\left( r\right) \mathbf{m\times }\left( \mathbf{%
m\times \hat{M}}_{p}\right)\,,  \label{Eq: LLG}
\end{equation}%
where $\gamma$ is the gyromagnetic ratio, $\alpha$ is the dimensionless Gilbert damping parameter, $\mathbf{m}$ and ${\mathbf{\hat{M}}}_p$ are the unit vectors denoting the local magnetization direction of the free layer and the uniform magnetization direction of the fixed layer respectively. The strength of the STT is characterized by an effective field $a_J(r)=a_J\mathcal{H}(R_c-r)$ with $\mathcal{H}(R_c-r)$ the Heaviside step function describing the confinement of the current in the NC-STO within a cylindrical region of radius $R_c$. The total effective magnetic field $\mathbf{H}_\mathrm{eff}$ is taken to be a superposition of the external field, the anisotropy field, the exchange field, and the demagnetization field, which can be expressed as
\begin{equation*}
\mathbf{H}_\mathrm{eff}=\mathbf{H}_{ext}+H_{a}m_{x}\boldsymbol{e}_{x}+\left(
2A_{ex}/M_{s}\right) \nabla ^{2}\mathbf{m-}4\pi M_{s}m_{z}\boldsymbol{e}_{z}\,,
\end{equation*}%
where $\mathbf{H}_{ext}$ is the uniform external field, $H_{a}$ the
magnitude of the anisotropy field, $A_{ex}$ the exchange stiffness, and the perpendicular-to-plane demagnetization field is reduced to the local form in the zero-thickness limit or thin-film approximation. To simplify our discussion, we shall restrict ourselves to a simple geometry where the the
magnetization of the fixed layer is lying along the $x$-axis\footnote{We note that in experiments, the application of a large external field will slightly tilt the magnetization of the fixed layer out of plane. We will ignore such out-of-plane components as they will be small and will primarily only lead to a small renormalization of the STT. Similarly, we are ignoring field-like STT as it would slightly renormalize the external magnetic field.}, i.e., $\mathbf{%
\hat{M}}_{p}=\boldsymbol{e}_{x}$, and both the magnetization of the free layer and the external magnetic field are varied within the $%
x-z$ plane, i.e., $\mathbf{m}=\cos \theta _{M}
\boldsymbol{e}_{x}+\sin \theta _{M}\boldsymbol{e}_{z}$ and $\mathbf{H}_{ext}=H_{ext}\left( \cos \theta _{H}
\boldsymbol{e}_{x}+\sin \theta _{H}\boldsymbol{e}_{z}\right)$.

At temperatures well below the Curie temperature, it is a good approximation to assume that the magnitude of the free layer magnetization $\mathbf{m}$ is conserved [as implied by Eq.~(\ref{Eq: LLG})]. This leaves only two independent components of $\mathbf{m}$, which can be expressed in terms of a single (space-dependent) complex spin wave variable $a({\mathbf r})=a\left[ \mathbf{m}({\mathbf r})\right] $, characterizing the amplitude and phase of spin waves~\cite{slavin2008ieeem,slavin2009ieee}. After performing a sequence of standard canonical transformations~\cite{Krasitskii1990JETP,slavin2008ieeem,Lvov}, one
arrives at the nonlinear spin wave dynamic equation~\cite{slonczewski1999jmmm,slavin2008ieeem,slavin2005prl}
\begin{equation}
\frac{\partial a}{\partial t}=-\mathrm{i}\left( \mathcal{\omega }_{r}-%
\mathcal{D}_{ex}\nabla ^{2}+\mathcal{N}_{f}\left\vert a\right\vert
^{2}\right) a+\mathcal{T}_{J}\left( r\right) \left( a-\left\vert
a\right\vert ^{2}a\right) -\mathcal{T}_{\alpha }\left( a+\mathcal{\kappa }%
\left\vert a\right\vert ^{2}a\right)\,,  \label{Eq: EOM-a}
\end{equation}%
where $\mathcal{\omega }_{r}=\gamma \sqrt{H_{int}\left( H_{int}+4\pi
M_{s}\cos ^{2}\theta _{M}-H_{a}\sin ^{2}\theta _{M}\right) }$ is the FMR
frequency of the uniform mode with $H_{int}$ the magnitude of the internal
magnetic field given by $\mathbf{H}_{int}=\mathbf{H}_{ext}+H_{a}m_{x}%
\boldsymbol{e}_{x}\mathbf{-}4\pi M_{s}m_{z}\boldsymbol{e}_{z}$ and $\cos
\theta _{M}=\mathbf{m\cdot }\boldsymbol{e}_{x}$, $\mathcal{T}_{J}\left(
r\right) =\gamma a_{J}\left( r\right) \cos \theta _{M}$ characterizes the
spin wave damping/pumping rate due to the STT, $\mathcal{D}_{ex}=\gamma
\left( A_{ex}/M_{s}\right) \left( \omega _{0}/\omega _{H}+\omega _{H}/\omega
_{0}\right) $ is the coefficient of the exchange spin wave dispersion with $%
\omega _{H}\equiv \gamma H_{int}$, $\mathcal{T}_{\alpha }=\alpha _{G}\left[
\omega _{H}+\left( \omega _{M}\cos ^{2}\theta _{M}-\omega _{A}\sin
^{2}\theta _{M}\right) /2\right] $ is the overall spin wave damping rate
with $\omega _{M}\equiv \gamma 4\pi M_{s}$ and $\omega _{A}\equiv \gamma
H_{a}$, $\mathcal{\kappa }$ measures the relative spin wave relaxation rates
of nonlinear and linear processes~\cite{slavin2008ieeem}, and $\mathcal{N}_{f}$ is the coefficient of the nonlinear spin wave frequency
shift which has been shown~\cite{slavin2008ieeem,Krivosik2010} to strongly depends on the out-of-plane angles of the external field and the
equilibrium magnetization, and may switch sign when $\mathbf{H}_{ext}$ varies
from in-plane to perpendicular-to-plane, as shown in Fig.~\ref%
{fig:Nf-v-thetaH}

\begin{figure}[tbp]
\includegraphics[width=.6\textwidth]{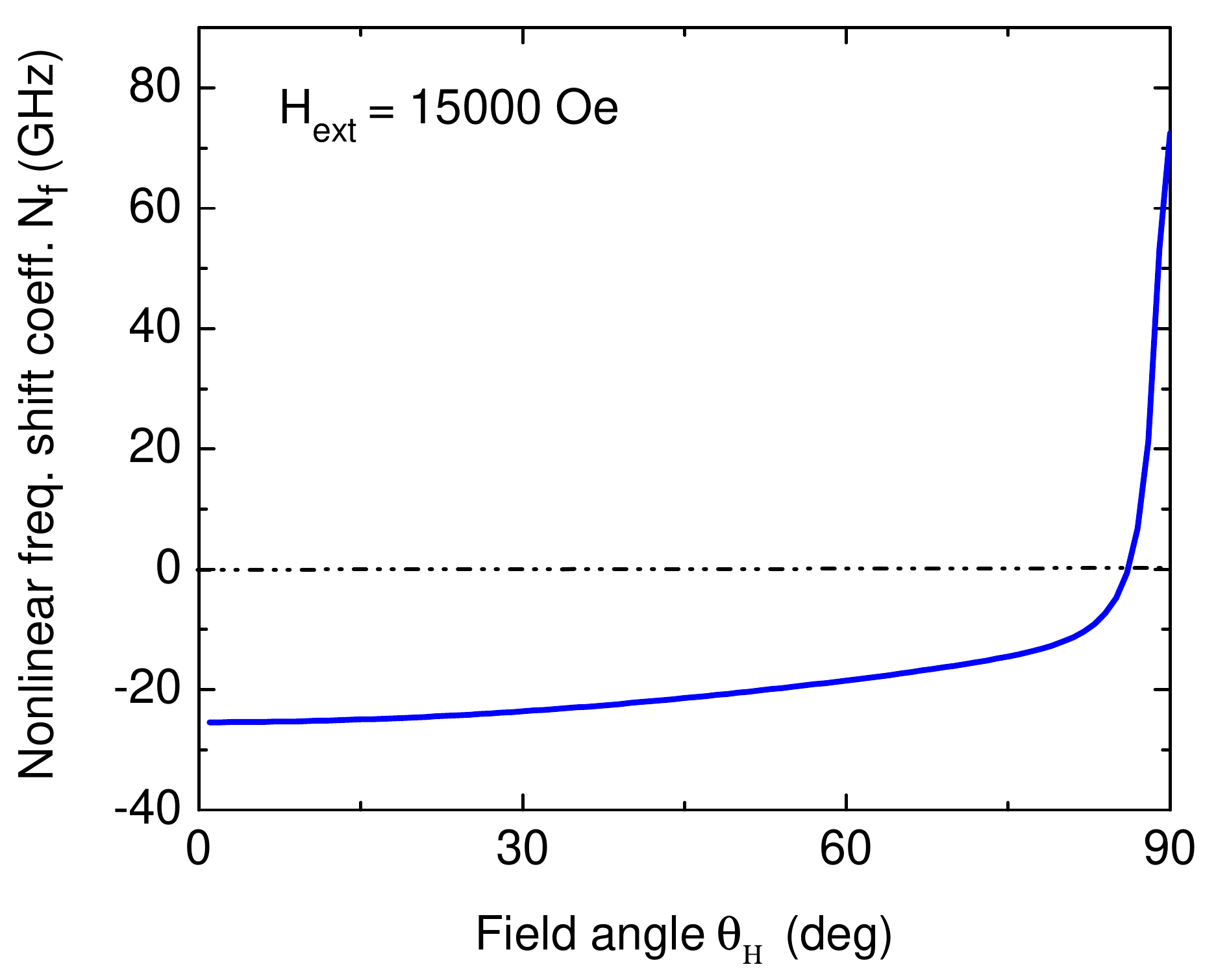}
\caption{Nonlinear frequency shift coefficient $\mathcal{N}_f$ as a
function of the out-of-plane field angle $\theta_{H}$. Other
parameters used here are $H_{ext}=15000$~Oe, $H_a=500$~Oe, $M_{S}=1000$~emu/cm$^3$,
$a_{J}=500$~Oe, the exchange length $l_{ex}=3$~nm [ $l_{ex}\equiv \sqrt{A_{ex}/2\pi M_s^2}$ ], $R_{c}=50$~nm, and radius
of the free layer $R_{F}=500$~nm.}
\label{fig:Nf-v-thetaH}
\end{figure}

Coupling between linear spin wave modes is induced by the cubic terms $%
\left\vert a\right\vert ^{2}a$ on the r.h.s. of Eq.~(\ref{Eq: EOM-a}), which
originates from the four-wave processes that conserve the number of spin
waves (other non-resonant wave processes can be eliminated by the
quasi-linear Krasitskii transformation~\cite{Lvov,Krasitskii1990JETP}). A complete set of orthonormal
linear spin wave eigenmodes $\left\{ d_{n}\left( r,t\right) \right\} $ (with
mode indices $n=1,2,..$) can be determined by solving the linearized, but including non-conservative torques, wave
equation
\begin{equation}
\frac{\partial d}{\partial t}=-\mathrm{i}\left( \mathcal{\omega }_{r}-%
\mathcal{D}_{ex}\nabla ^{2}\right) d-\mathcal{T}_{\alpha }d+\mathcal{T}%
_{J}\left( r\right) d\,.  \label{Eq: EOM-d}
\end{equation}%
The solution of Eq.~(\ref{Eq: EOM-d}) was first obtained by Slonczewski for
a perpendicularly magnetized thin film~\cite{slonczewski1999jmmm},
and later by Slavin \emph{et al.}~\cite{slavin2005prl} for more general
cases [see Appendix A for an outline of the solution]. We then may expand
the general spin wave mode in this basis, i.e.,
\begin{equation}
a\left( r,t\right) =\sum_{n}A_{n}\left( t\right) \left[ u_{n}\left( r\right)
e^{-\mathrm{i}\omega _{n}t}\right]\,,  \label{Eq: a-d_n_expan}
\end{equation}%
where $A_{n}\left( t\right) $ are complex coefficients that describe the composition of a given spin wave
mode in terms of the linear eigenmodes, and we have separated the spatial
and temporal components of the eigenmodes as $d_{n}\left( r,t\right) =u_{n}\left(
r\right) e^{-\mathrm{i}\omega _{n}t}$, where $\omega_n$ are the
complex eigenfrequencies of the linearized modes. We will assume that the system is operating close to, but above, the critical current for auto-oscillations so that the imaginary part of $\omega_n$ is small and can be ignored. Note that since the experimentally observed time evolution of coupled modes~\cite{Muduli12PRB_T-dep-STO,muduli2012prl,dumas2013prl} is usually slower than the periods of the eigenmodes $2\pi/\mathfrak{R}_{e}(\omega_n)$, it is reasonable to assume the characteristic time scale of the variation of $A_n$ is greater than the periods of eigenmodes of interests. By placing the expansion~(\ref{Eq: a-d_n_expan}) into Eq.~(\ref{Eq: EOM-a}),
projecting with $d_{i}^{\ast }$, and integrating the resulting equation over a time interval spanning several eigenmode periods for which the slowly varying amplitude of the $i$-th mode $A_{i}\left(t\right)$ remains approximately constant~\cite{Zhang16jmmm}, we arrive at the following rate equation for $A_{i}\left(t\right)$:
\begin{equation}
\frac{dA_{i}}{dt}=-\sum_{l,m,n}\int\limits_{0}^{R_{F}/R_{c}}\mathrm{d}^{2}%
\mathbf{r}^{\prime }\left[ \mathrm{i}\mathcal{N}_{f}+\mathcal{\kappa T}%
_{\alpha }+\mathcal{T}_{J}\left( r^{\prime }\right) \right] u_{i}^{\ast
}\left( r^{\prime }\right) u_{m}^{\ast }\left( r^{\prime }\right)
u_{l}\left( r^{\prime }\right) u_{n}\left( r^{\prime }\right) A_{m}^{\ast
}A_{l}A_{n}\delta _{\omega _{i}+\omega _{m},\omega _{l}+\omega _{m}}\,,
\label{Eq:EOM-A}
\end{equation}%
where $r^{\prime }\equiv r/R_{c}$ is the dimensionless radial distance, and we
used the completeness relation
$\left( 2\pi \right) ^{-1}\int \mathrm{d}te^
{-{\mathrm i}\left(\omega_n-\omega_{n'}\right)t}
=\delta
_{n,n^{\prime }}$.

As a minimal model to capture the essential physics underlying the coupling
of linear spin wave modes, let us focus on interactions between the two
nondegenerate lowest lying modes, namely, $A_{1}$ and $A_{2}$ with
eigenenergies of $\mathcal{\omega }_1$ and $\mathcal{\omega }_{2}$ [assuming ${\omega }_1<{\omega }_{2}$]. By
imposing energy conservation on the four-wave processes, three
types of terms enter the mode equation: 1) the self-energy terms $%
A_{1}^{\ast }A_{1}A_{1}$ and $A_{2}^{\ast }A_{2}A_{2}$, 2) the mutual energy
transfer terms $A_{1}^{\ast }A_{1}A_{2}$ and $A_{2}^{\ast }A_{1}A_{2}$, and
3) the terms $A_{m}^{\ast }A_{n}A_{2}$ [with $m>n>2$] which correspond to the
four-wave processes of $a_{m}^{\ast }a_{2}^{\ast }a_{n}a_{1}$. While the first two types of terms give rise to the nonlinear coupling terms~\cite{deAguiar2007prb,muduli2012prl,Zhang16jmmm}, the third type of terms will lead to a linear
coupling between modes $1$ and $2$ when the higher-energy modes are thermally excited, which is usually the case for a NC-STO operated at room temperature. In this case, we can replace $A_{m}^{\ast }$ and $A_{n}$ with thermal occupation numbers of the magnon modes $m$ and $n$ by taking the trace of the density matrix over
magnon Fock space with $m$ and $n$ magnons. However, keeping in mind that the
amplitude $A_m$ of a magnon corresponds to a reduction of the total magnetization of $\sim n_B(\omega_m)g\mu_B/(M_SV)=n_B(\omega_m)/(NS)$, where $V$ is the volume and $NS$ the total atomic spin of the free layer,
we scale the occupation numbers appropriately, i.e., $\tilde n_{B}\left( \omega _{m}\right) \equiv \left[
n_{B}\left( \omega _{m}\right) +1\right] /(NS)\simeq n_{B}\left( \omega
_{m}\right) /(NS)$ and $\tilde n_{B}\left( \omega _{n}\right) \equiv
n_{B}\left( \omega _{n}\right) /(NS)$ respectively,
where $n_{B}\left( \omega _{n}\right)
=1/\left( e^{\hbar \omega _{n}/k_{B}T}-1\right)$. After
collecting all relevant terms, we arrive at a set of rate equations describing the coupled dynamics of the two modes
\begin{eqnarray}
\frac{dA_{1}(t)}{dt} &=&-\mathrm{i}\left( \eta _{1,1}|A_{1}|^{2}+\eta
_{1,2}|A_{2}|^{2}\right) A_{1}-\Gamma _{G,1}\left(
P_{1,1}|A_{1}|^{2}+P_{1,2}|A_{2}|^{2}\right) A_{1}  \notag \\
&&-\Gamma _{J}\left( Q_{1,1}|A_{1}|^{2}+Q_{1,2}|A_{2}|^{2}\right)
A_{1}-R_{1,2}(T)A_{2}  \label{Eq:EOM-A1} \\
\frac{dA_{2}(t)}{dt} &=&-\mathrm{i}\left( \eta _{2,1}|A_{1}|^{2}+\eta
_{2,2}|A_{2}|^{2}\right) A_{2}-\Gamma _{G,2}\left(
P_{2,1}|A_{1}|^{2}+P_{2,2}|A_{2}|^{2}\right) A_{2}  \notag \\
&&-\Gamma _{J}\left( Q_{2,1}|A_{1}|^{2}+Q_{2,2}|A_{2}|^{2}\right)
A_{2}-R_{2,1}(T)A_{2}\,,  \label{Eq:EOM-A2}
\end{eqnarray}%
where $\Gamma _{G,i}=\alpha \omega _{i}$ ($i,j=1$ or $2$), $\Gamma
_{J}=\gamma a_{J}\cos \theta_{M}, $ $\eta _{i,j}\equiv \mathcal{N}%
_{f}\int\limits_{0}^{R_{F}/R_{c}}\mathrm{d}^{2}\mathbf{r}^{\prime
}\left\vert u_{i}u_{j}\right\vert ^{2}$ are the nonlinear frequency shift
coefficients, $P_{i,j}=\kappa \int\limits_{0}^{R_{F}/R_{c}}\mathrm{d}^{2}%
\mathbf{r}^{\prime }\left\vert u_{i}u_{j}\right\vert ^{2}$ are the nonlinear
damping coefficients, $Q_{i,j}=\int\limits_{0}^{1}\mathrm{d}^{2}\mathbf{r}%
^{\prime }\left\vert u_{i}u_{j}\right\vert ^{2}$ are the nonlinear
coefficients associated with STT term, and the linear mode coupling
coefficients $R_{i,j}$ are
\begin{eqnarray}
R_{i,j}\left( \mathbf{H}_{ext},T\right) =\sum\limits_{n,m}
&&\int\limits_{0}^{R_{F}/R_{c}}\mathrm{d}^{2}\mathbf{r}^{\prime }u_{i}^{\ast
}\left( r^{\prime }\right) u_{m}^{\ast }\left( r^{\prime }\right)
u_{j}\left( r^{\prime }\right) u_{n}\left( r^{\prime }\right) \left[ \mathrm{%
i}\mathcal{N}_{f}+\mathcal{\kappa T}_{\alpha }+\mathcal{T}_{J}\left(
r^{\prime }\right) \right]  \notag \\
&&\times \tilde n_{B}(\omega _{n})\tilde n_{B}(\omega _{m})\delta
_{\omega _{n}+\omega _{2},\omega _{m}+\omega _{1}}\,.  \label{Eq:R_ij}\nonumber\\
\end{eqnarray}
Note that the temperature dependence of the $R_{i,j}$ enters through the
magnon thermal distribution functions $\tilde n_{B}$.

\begin{figure}[tbp]
\includegraphics[width=.6\textwidth]{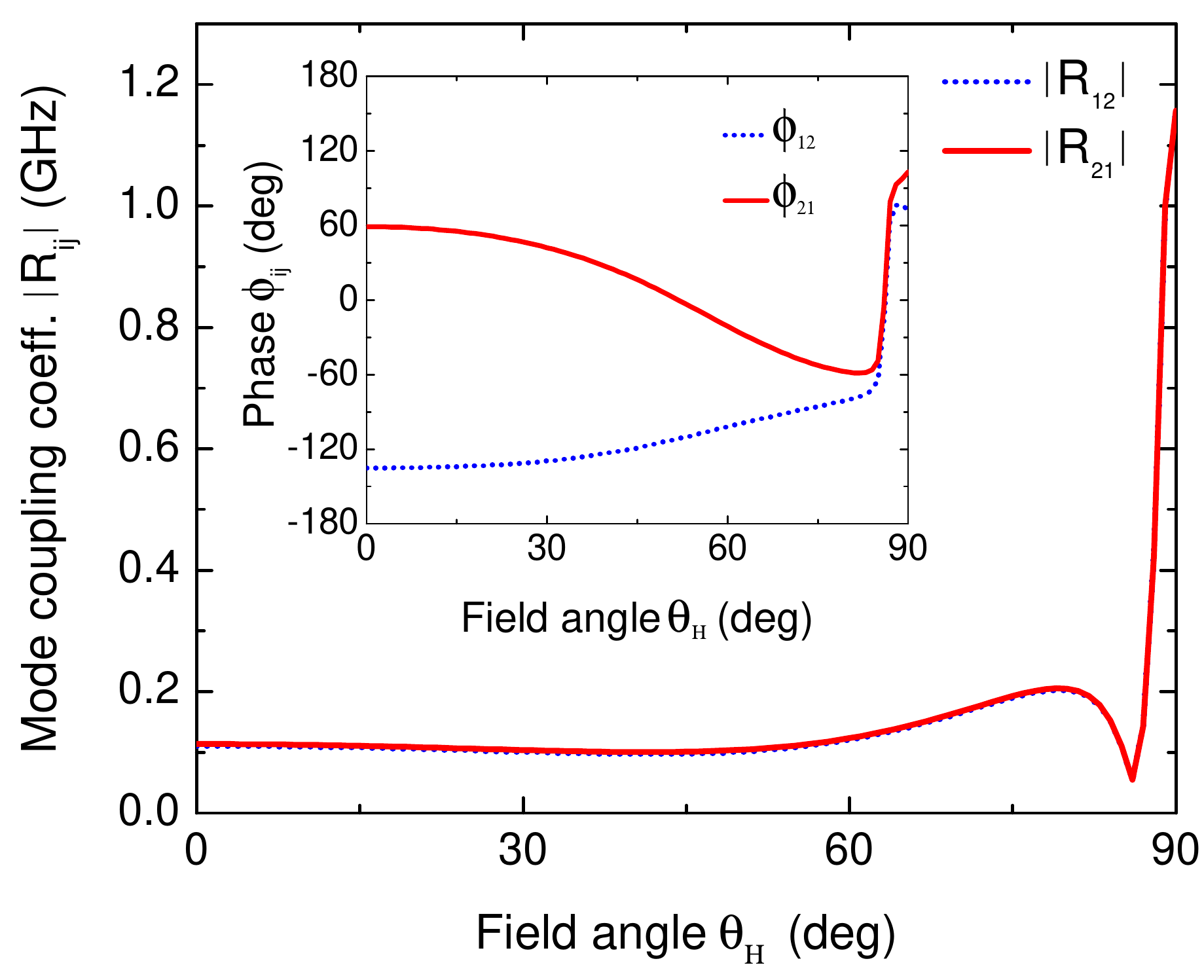}
\caption{Magnitude of the mode coupling coefficients $R_{1,2}$ and $R_{2,1}$
as functions of the angle $\protect\theta _{H}$ between the film plane and
the external magnetic field of fixed strength $15000$~Oe at room
temperature. In the inset of the figure, we show the phases of $R_{1,2}$ and
$R_{2,1}$ as functions of $\protect\theta_{H}$. Other parameters used here
are $\alpha=0.05$, $M_{s}=1000$~emu/cm$^3$, $a_{J}=500$~Oe, $%
R_{c}=50$~nm, the exchange length $l_{ex}=3$~nm and radius of the free
layer $R_{F}=500$~nm.}
\label{fig:R-v-thetaH}
\end{figure}

Equipped with Eq.~(\ref{Eq:R_ij}), we are now in a position to investigate the
dependence of the mode coupling coefficient $R_{1,2}$ on external experimental conditions as
well as intrinsic magnetic properties. In Fig.~\ref{fig:R-v-thetaH}, we show the magnitudes
and phases of the linear mode coupling coefficients $R_{1,2}$ and $R_{2,1}$
as functions of the out-of-plane field angle $\theta_H$. We note that the magnitudes of coupling
coefficients are about sub-GHz for an applied external field of $15000$~Oe.
This justifies our assumption that the time evolution of $A_n(t)$ ($n=1,2$)
is slow compared to the fast dynamics of the magnetization. Also,
the magnitude of the coefficients $R_{i,j}$ approaches a global minimum at a field angle of $%
\theta_H=86^{\circ}$, which coincides with the field angle at which the
non-linear spin wave frequency shift coefficient $\mathcal{N}_f$ changes
sign, as demonstrated in Fig.~\ref{fig:Nf-v-thetaH}. In addition, the phase difference increases as the field angle moves away from the zero-crossing point of the nonlinear frequency shift. The link between the
linear mode coupling and the nonlinear spin wave frequency shift may suggest a way
to control and manipulate the mode coupling.

\begin{figure}[tbp]
\includegraphics[width=.6
\textwidth]{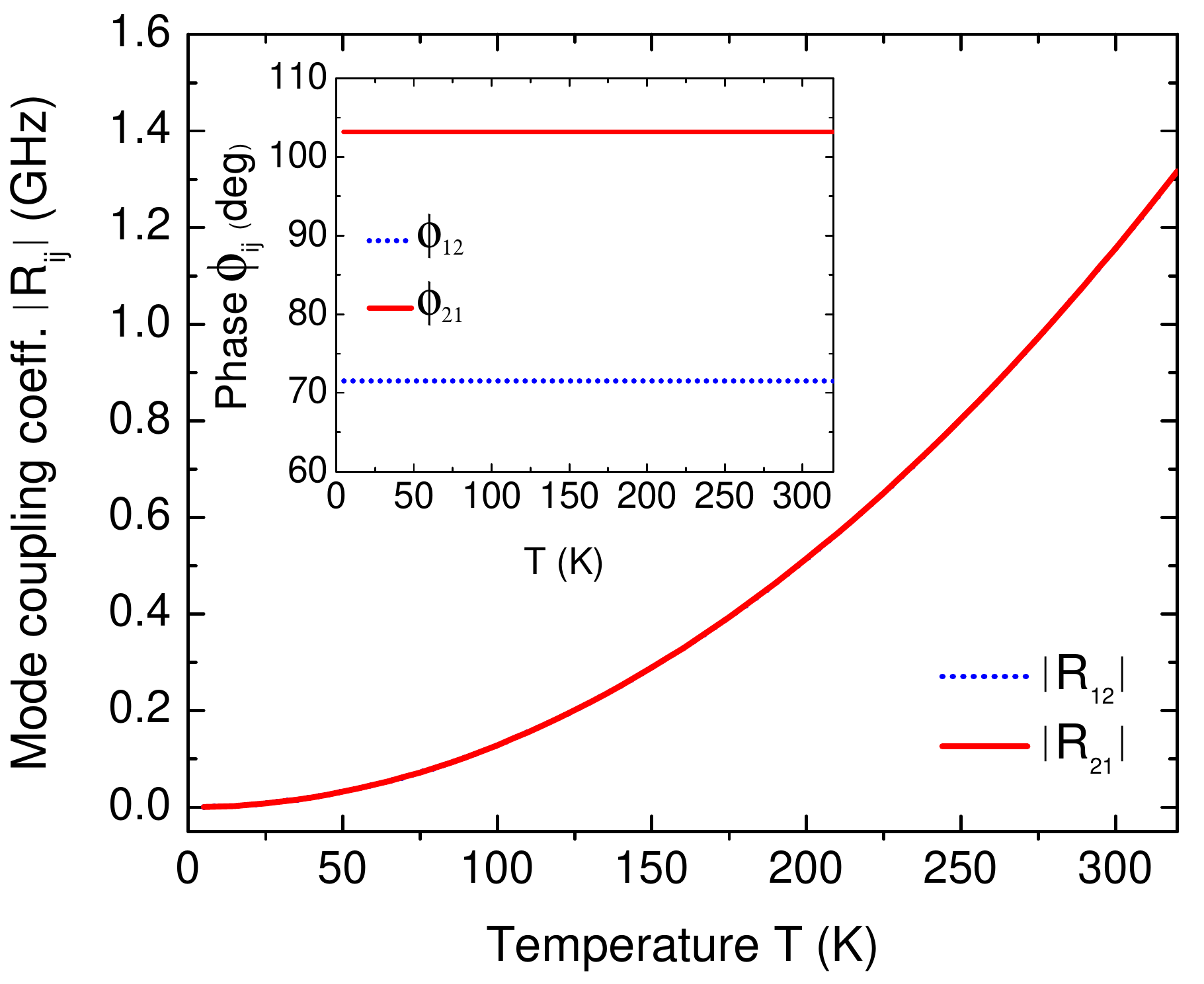}
\caption{Magnitude of the mode coupling coefficients $R_{1,2}$ and $R_{2,1}$
as functions of temperature $T$ with an external magnetic field $H_{ext}$ of
$15000$~Oe applied perpendicular to the plane of the layer. In the inset of the
figure, we show the phase of $R_{1,2}$ and $R_{2,1}$ as functions of $T$.
Other parameters: $\protect\alpha _{G}=0.05$, $M_{S}=1000$~emu/cm$^3$, $%
a_{J}=500$~Oe, $R_{c}=50$~nm, the exchange length $l_{ex}=3$~nm, and
radius of the free layer $R_{F}=500$~nm. }
\label{fig:R-v-T}
\end{figure}

In Fig.~\ref{fig:R-v-T}, we show the temperature dependence of the mode
coupling coefficients $R_{1,2}$ and $R_{2,1}$ for $H_{ext}=15000$~Oe at several out-of-plane field angles. We see that the magnitudes of both $R_{1,2}$ and $R_{2,1}$ increase algebraically with temperature. This can be understood by recalling that the linear mode coupling stems from interaction\textbf{s} between the two dominant modes and thermally occupied magnons via the four-magnon scattering processes. The density of thermal magnons increases at elevated temperatures, which gives rise to more scattering space
that contributes to the linear mode coupling. This temperature dependence, as we will show later, implies that temperature alone can change the manifold of
the system's dynamics and
leads to thermally-induced mode-hopping~\cite{Iacocca2014prb} consistent with experimental observations~\cite{Muduli12PRB_T-dep-STO}. The phases of the coupling
coefficients, however, are insensitive to temperature, as shown in the inset
of Fig.~\ref{fig:R-v-T}, since we have implicitly made the random phase
approximation whereby the thermal magnons of different wave-vectors are
taken to be incoherent.

\section{Phase portrait of the mode equations}

Although the energy of an individual mode varies instantaneously through
pumping and damping, we may assume that the total energy of the two dominant
modes $1$ and $2$ are approximately conserved on
time scales much longer than the periods of the eigenmodes, i.e.,
\begin{equation}
\omega _{1}\left\vert A_{1}\right\vert ^{2}+\omega _{2}\left\vert
A_{2}\right\vert ^{2}=\omega\,.  \label{Eq: en-consv-1&2}
\end{equation}%
where $\omega $ is a constant, the value of which indicates the total energy in the two-mode subsystem. Substitution of $A_{i}\left( t\right)
=K_{i}\left( t\right) e^{\mathrm{i}\phi _{i}\left( t\right) }$ [$i=1,2$]
into Eq.~(\ref{Eq: en-consv-1&2}) immediately reveals the relation between the amplitudes of the two modes, which can be captured by following
transformation~\cite{beri2008prl}: $\sqrt{\omega _{1}}K_{1}=\sqrt{\omega }\cos \left( \frac{%
\theta }{2}+\frac{\pi }{4}\right) $ and $\sqrt{\omega _{2}}K_{2}=\sqrt{%
\omega }\sin \left( \frac{\theta }{2}+\frac{\pi }{4}\right) $, where the
variable $\theta $ characterizes the relative magnitude of the two mode
amplitudes. Placing these transformations into the mode equations (\ref%
{Eq:EOM-A1}) and (\ref{Eq:EOM-A2}) and separating the real and imaginary
parts of the resulting equations, we arrive at a set of coupled dynamic
equations for two real variables, i.e.,
\begin{eqnarray}
\dot{\theta} &=&\omega \cos \theta \left[ \left( \Gamma _{G,1}P_{1,1}-\Gamma
_{G,2}P_{2,1}\right) \left( \frac{1-\sin \theta }{2\omega _{1}}\right)
-\left( \Gamma _{G,2}P_{2,2}-\Gamma _{G,1}P_{1,2}\right) \left( \frac{1+\sin
\theta }{2\omega _{2}}\right) \right]  \notag \\
&&+\omega \Gamma _{J}\cos \theta \left[ \left( Q_{1,1}-Q_{2,1}\right) \left(
\frac{1-\sin \theta }{2\omega _{1}}\right) -\left( Q_{2,2}-Q_{1,2}\right)
\left( \frac{1+\sin \theta }{2\omega _{2}}\right) \right]  \notag \\
&&+\mathfrak{R}_{e}\left( R_{2,1}e^{-i\psi }\right) \sqrt{\frac{\omega _{2}}{%
\omega _{1}}}\left( 1-\sin \theta \right) -\mathfrak{R}_{e}\left(
R_{1,2}e^{i\psi }\right) \sqrt{\frac{\omega _{1}}{\omega _{2}}}\left( 1+\sin
\theta \right)  \label{Eq: EOM-theta}
\end{eqnarray}%
and
\begin{eqnarray}
\dot{\psi} &=&\omega \left[ \left( \eta _{1,1}-\eta _{2,1}\right) \left(
\frac{1-\sin \theta }{2\omega _{1}}\right) -\left( \eta _{2,2}-\eta
_{1,2}\right) \left( \frac{1+\sin \theta }{2\omega _{2}}\right) \right]
\notag \\
&&+\mathfrak{I}_{m}\left( R_{2,1}e^{-i\psi }\right) \sec \theta \sqrt{\frac{%
\omega _{2}}{\omega _{1}}}\left( 1-\sin \theta \right) -\mathfrak{I}%
_{m}\left( R_{1,2}e^{i\psi }\right) \sec \theta \sqrt{\frac{\omega _{1}}{%
\omega _{2}}}\left( 1+\sin \theta \right)\,,  \label{Eq: EOM-psi}
\end{eqnarray}%
where $\psi \equiv \phi _{2}-\phi _{1}\,$is the phase difference of the two
modes. As we can see, the original mode equations, which involve four
independent dynamical variables, have been mapped onto a 2-dimensional phase
space, similar to those describing the dynamics of a ring laser with backscattering~\cite{beri2008prl,vanderSande}.

In a previous work~\cite{Iacocca2014prb}, a similar set of
the coupled effective mode equations was solved for a given external field and temperature, with the various
coefficients being treated as phenomenological parameters. Notably, the coefficients of the linear coupling terms were assumed to be complex conjugates which is manifestly \emph{not} generally true (see Fig.~\ref{fig:R-v-thetaH} and Fig.~\ref{fig:R-v-T}). Now that we have
derived these coefficients as functions of the external field and
temperature, we are in a position to further investigate the evolution of
the two modes in the phase space spanned by $\theta $ and $\psi $ for
varying field angles and temperatures. We note that because the system of equations~(\ref{Eq: EOM-theta}) and (\ref{Eq: EOM-psi}) is invariant under the transformations $\theta\rightarrow \pm \pi - \theta$ and $\psi \rightarrow \psi \pm \pi$, in principle it suffices to show the phase space for $-\frac{\pi}{2}<\theta<\frac{\pi}{2}$; however, we choose to show the phase space in the extended zone of  $-\pi<\theta<\pi$ in order to avoid confusing overlay of trajectories corresponding to in-phase and out-of-phase solutions with crossing of trajectories, which is prohibited by the uniqueness of solutions for dynamical systems that evolve smoothly. Note that this extended scheme will depict the singularities at $\theta=\pm \frac{\pi}{2}$ given by Eq.~(\ref{Eq: EOM-psi}).

\begin{figure}[h]
\centering \includegraphics[width=5.in, trim=0.1in 0.1in 0in 0, clip]{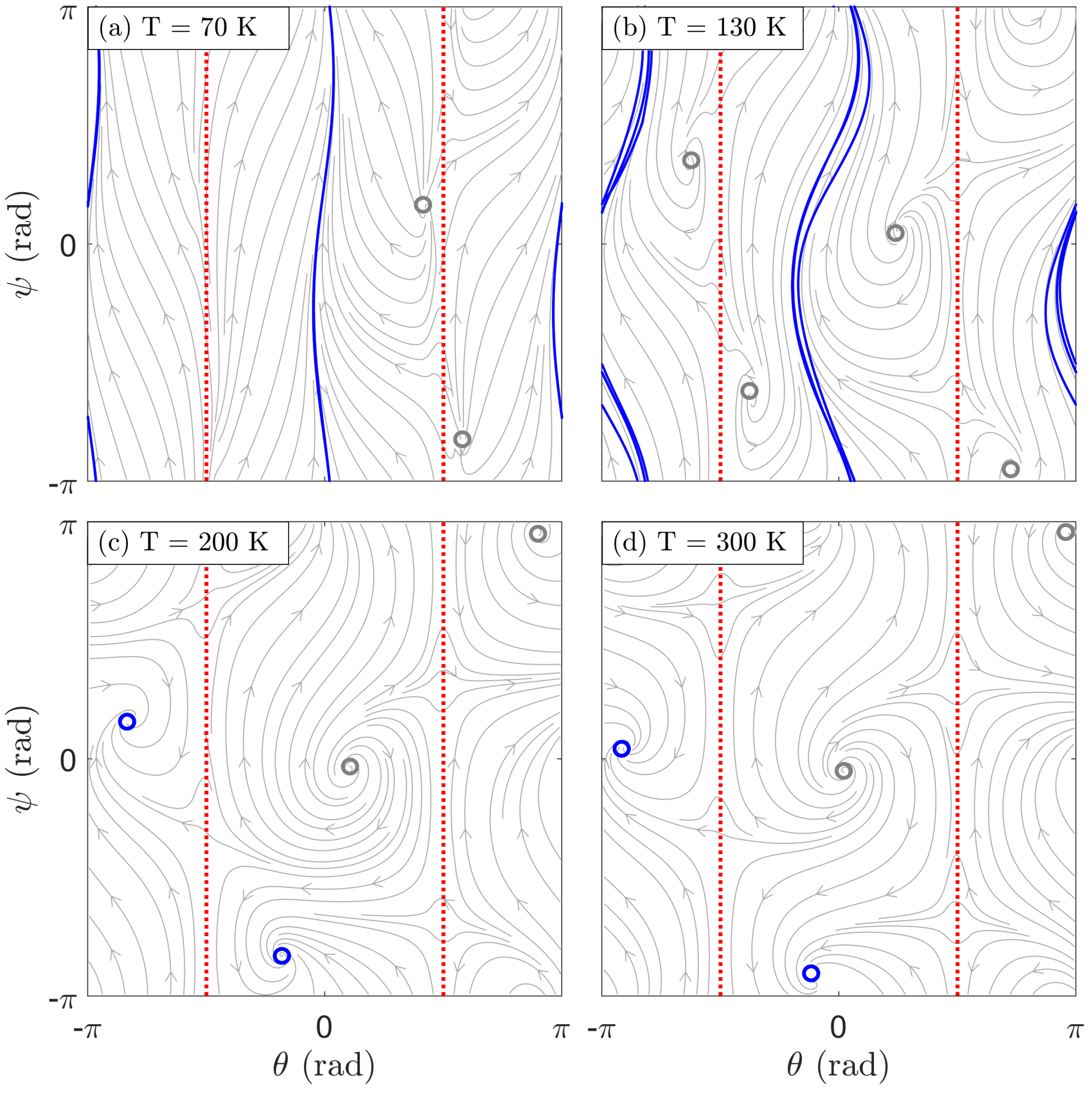}
\caption{Portrait of the dynamics of two coupled modes in the phase space
spanned by $\protect\theta$ and $\protect\psi$ for an external field of %
15000~Oe and $\protect\theta_H=82^{\circ}$ at (a) $T=70$~K (b) $T=130$~K, (c) $T=200$~K, and (d) $T=300$~K. The vertical red dotted lines denote the singularities at $\theta=\pm \frac{\pi}{2}$. As the mode coupling magnitude increases with temperature, the trajectories (blue lines) established by unstable fixed points (gray circles) transition into stable fixed points (blue circles). In all cases, the features are even indicating in-phase and out-of-phase synchronized modes. }
\label{fig:2d-phase-v-T}
\end{figure}

In Fig.~\ref{fig:2d-phase-v-T}, we show the phase portraits of mode dynamics for $\omega=1$~GHz at several different temperatures with a given external field of magnitude 15000~Oe and angle $\theta_H=82^{\circ}$. In this case, the coupling phase is constant and only the magnitude increases with temperature, similar to the perpendicular-to-plane case shown in Fig.~\ref{fig:R-v-T}.
Well below room temperature ($T=300~K$), pairs of unstable fixed points (solid gray circles) are present as shown in Fig.~\ref{fig:2d-phase-v-T}
(a) and (b), which are accompanied by steady state trajectories (solid blue lines) representing
coexistence of the two modes with periodic mutual energy transfer due to the mode coupling. As temperature increases, some of the unstable fixed points are converted into stable ones due to the increased strength of the linear mode coupling (as shown in Fig.~\ref{fig:2d-phase-v-T}(c)).
When the temperature is further increased, the linear mode coupling will dominate and the fixed points have to approach $\theta=\pm\pi$ or $\theta=0$, as indicated by Fig.~\ref{fig:2d-phase-v-T}(d).

Physically, stable fixed points correspond to phase-locked or synchronized modes, i.e., the differences in both phase and energy of the two modes remain constant in time. The existence of equal numbers of stable and unstable fixed points across the singularities (red dotted lines) is consistent with an in-phase/out-of-phase synchronization. These results demonstrate
that the phase portrait of the multi-mode dynamics in a NC-STO can be strongly affected by temperature and can lead to mode transitions, in agreement with experimental observations of temperature-dependent mode transitions above room temperature~\cite{Muduli12PRB_T-dep-STO}. We
note that in the calculation of the phase portraits, the temperature dependence only enters through the linear mode coupling coefficients $R_{i,j}$ in Eqs.~\eqref{Eq: EOM-theta} and \eqref{Eq: EOM-psi}. Inclusion of temperature as a stochastic field will merely blur the phase portraits and potentially lead to mode transitions across saddle points on the stable manifolds, leading to mode-hopping~\cite{muduli2012prl,Iacocca2014prb}, but not change the orbits or manifolds themselves.

\begin{figure}[h]
\centering \includegraphics[width=5.in, trim=0.1in 0.1in 0in 0, clip]{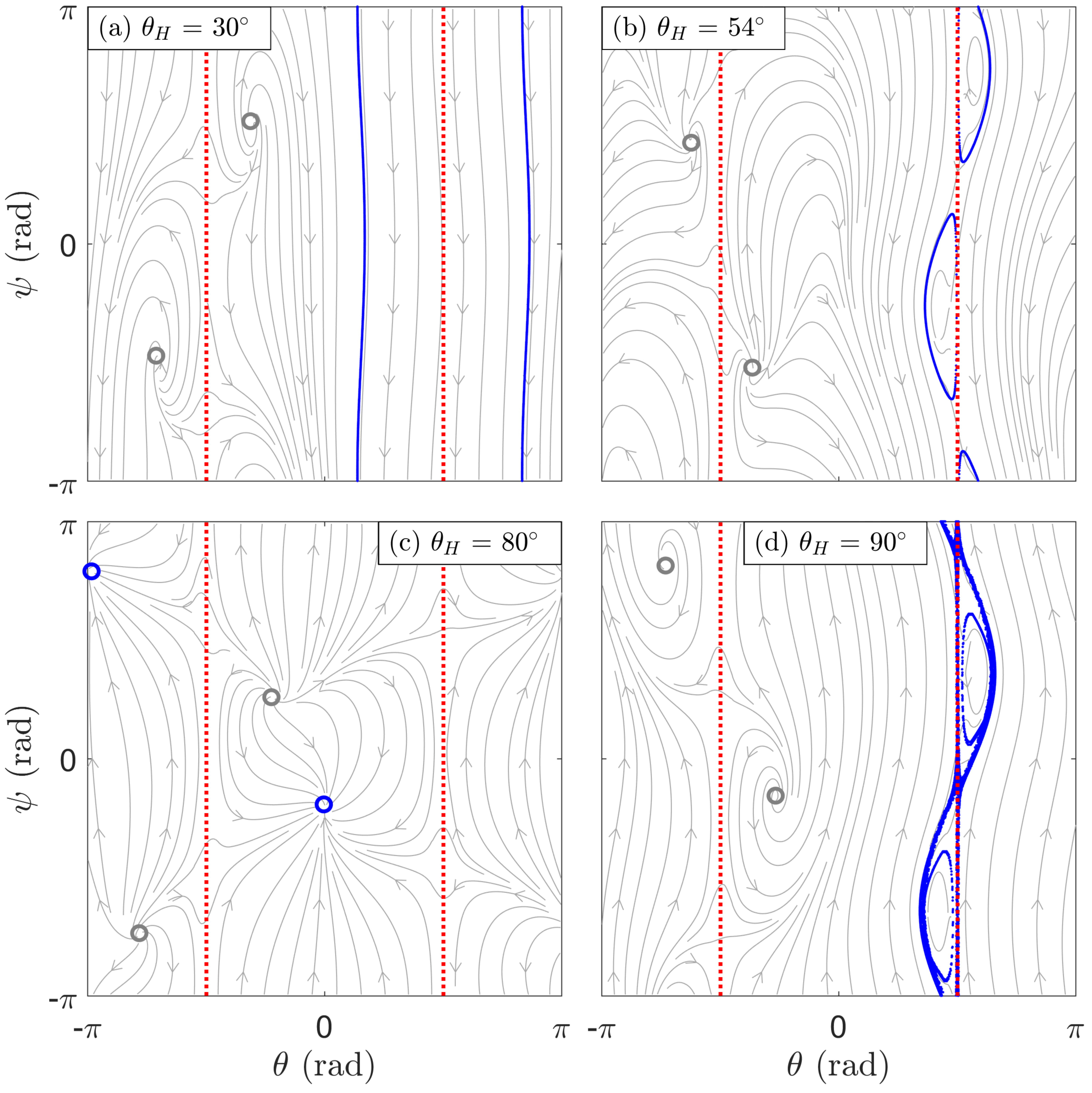}
\caption{Portrait of the dynamics of two coupled modes in the phase space
spanned by $\protect\theta$ and $\protect\psi$ for an external field of $%
15000$~Oe and $T = 200$~K at (a) $\theta_H=30^\circ$, (b) $\theta_H=54^\circ$, (c) $\theta_H=80^\circ$, and (d) $\theta_H=90^\circ$. (a) Unstable fixed points (gray circles) define trajectories (blue lines) for small external field angles. (b) Closed orbits (blue) are observed when the coupling angles are similar. (c) Stable and unstable fixed points (blue and gray circles) are observed at the local coupling strength maximum and identical coupling angles. (d) Closed orbits and heterloclinic-like orbits are observed at the high-symmetry condition of a perpendicularly magnetized sample. }
\label{fig:2d-phase-v-thetaH}
\end{figure}

More complex behavior\sout{s} are found when the external field angle $\theta_H$ is varied at a fixed temperature $T=200$~K (as shown in Fig.~\ref{fig:2d-phase-v-thetaH}), in which case both the magnitude and phase of the linear mode coupling coefficients $R_{i,j}$ vary with $\theta_H$.
We consider four regimes as the field angle increases: (\emph{i}) weak coupling magnitude and large phase difference, (\emph{ii}) weak coupling magnitude and small phase difference; (\emph{iii}) local coupling magnitude maximum and small phase difference; and (\emph{iv}) strong coupling magnitude and identical phases. Exemplary phase portraits of these regimes are shown in Fig.~\ref{fig:2d-phase-v-thetaH} as the angle is varied from in-plane to perpendicular-to-plane. For case (\emph{i}) [Fig.~\ref{fig:2d-phase-v-thetaH} (a)], the phase portrait exhibits two unstable fixed points (spirals) and stable steady state trajectories, similar to the phase portraits well below room temperature as shown in Fig.~\ref{fig:2d-phase-v-T}(a). For case (\emph{ii}) shown in Fig.~\ref{fig:2d-phase-v-thetaH}(b), the trajectories get pulled around to form the large elliptical closed orbits which imply self-sustained periodic oscillation of the coupled subsystem. For case (\emph{iii}) [Fig.~\ref{fig:2d-phase-v-thetaH}(c)], the two closed orbits collapse to two stable fixed points as the system reaches a local maximum of linear mode  coupling strength (see Fig.~\ref{fig:R-v-thetaH}). Finally, for (\emph{iv}), the phase portrait is again dominated by closed orbits [Fig.~\ref{fig:2d-phase-v-thetaH} (d)] accompanied by heteroclinic-like orbits and chaotic dynamics near the singular line $\theta=\pi/2$. This observation is consistent with the poor spectral content of STT-driven excitations in perpendicularly magnetized STOs~\cite{Dumas2014} despite the fact that such a high-symmetry case theoretically favors a Slonczewski mode in a single-mode approximation.

\begin{figure}[h]
\centering \includegraphics[width=5.in, trim=0.1in 0.1in 0in 0, clip]{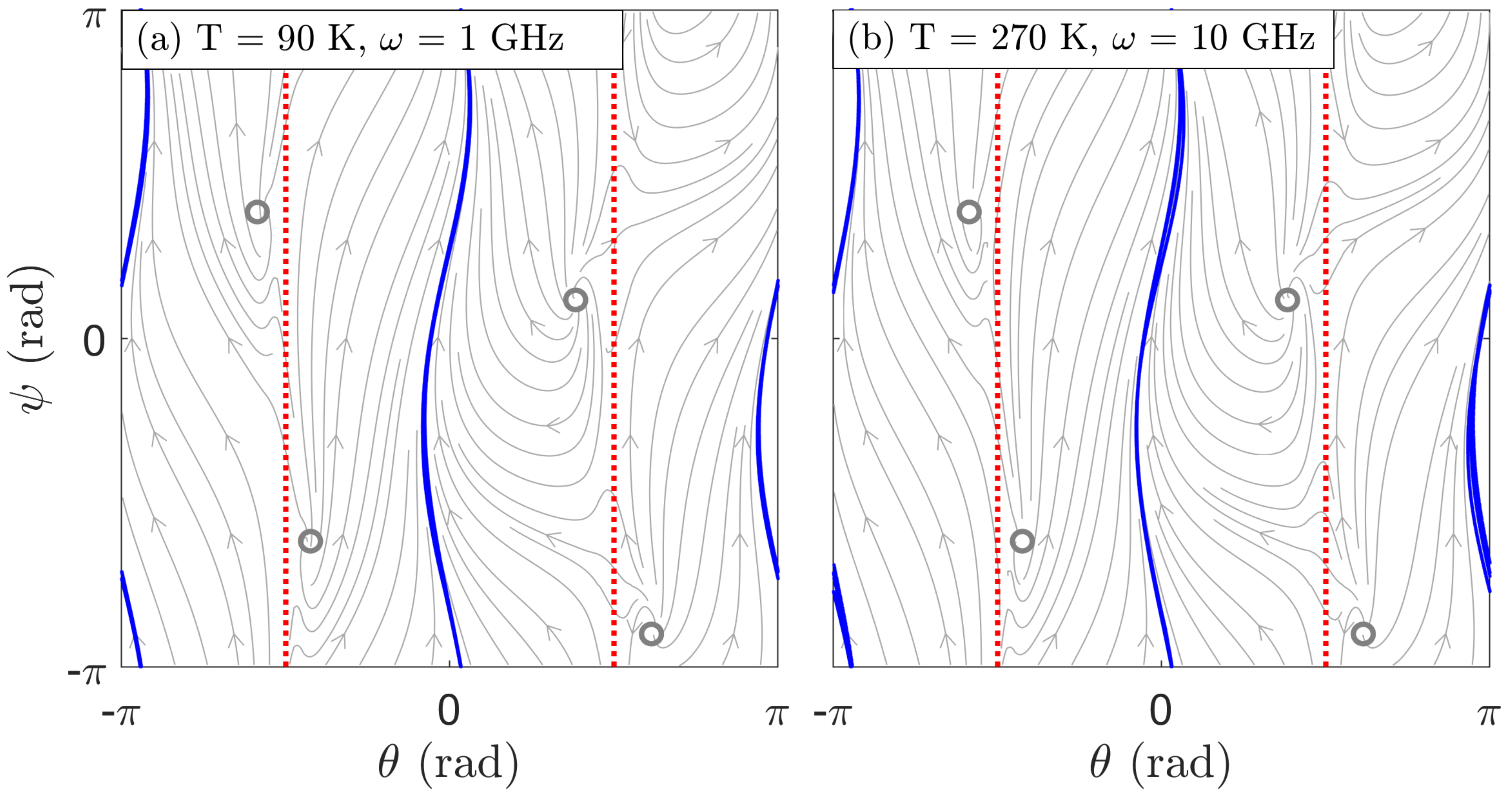}
\caption{Portrait of the dynamics of two coupled modes in the phase space
spanned by $\protect\theta$ and $\protect\psi$ for an external field of 15000~Oe and $\protect\theta_H=82^{\circ}$ at (a) $T=90$~K, $\omega=1$~GHz and (b) $T=270$~K, $\omega=10$~GHz. Both phase portraits are equivalent. }
\label{fig:2d-phase-v-omega}
\end{figure}

In addition to the field and temperature dependence of the coupled mode dynamics, another observation is that increasing/decreasing the amount of energy in the two-mode subsystem (characterized by $\omega$) is equivalent to reducing/enhancing the strength of the linear coupling coefficients $R_{i,j}$. This can be seen if one divides both sides of Eqs.~(\ref{Eq: EOM-theta}) and (\ref{Eq: EOM-psi}) by $\omega$, and the resulting equations indicate the steady state solutions of the system only rely on the ratio $|R_{i,j}|/\omega$. To illustrate this effect, we show in Fig.~\ref{fig:2d-phase-v-omega} the nearly identical phase portraits at a fixed field angle of $\theta_H=82^{\circ}$ for two different sets of parameters: (a) temperatures and (a) $T=90$~K, $\omega=1$~GHz and (b) $T=270$~K, $\omega=10$~GHz. This is consistent with the fact that the strength of linear mode coupling coefficient for $\theta_H=82^{\circ}$ at $T=270$~K is about 10 times as large as that for $T=90$~K, according to Fig.~\ref{fig:R-v-T}.

\begin{figure}[h]
\centering \includegraphics[width=5.in, trim=0.1in 0.1in 0.1in 0, clip]{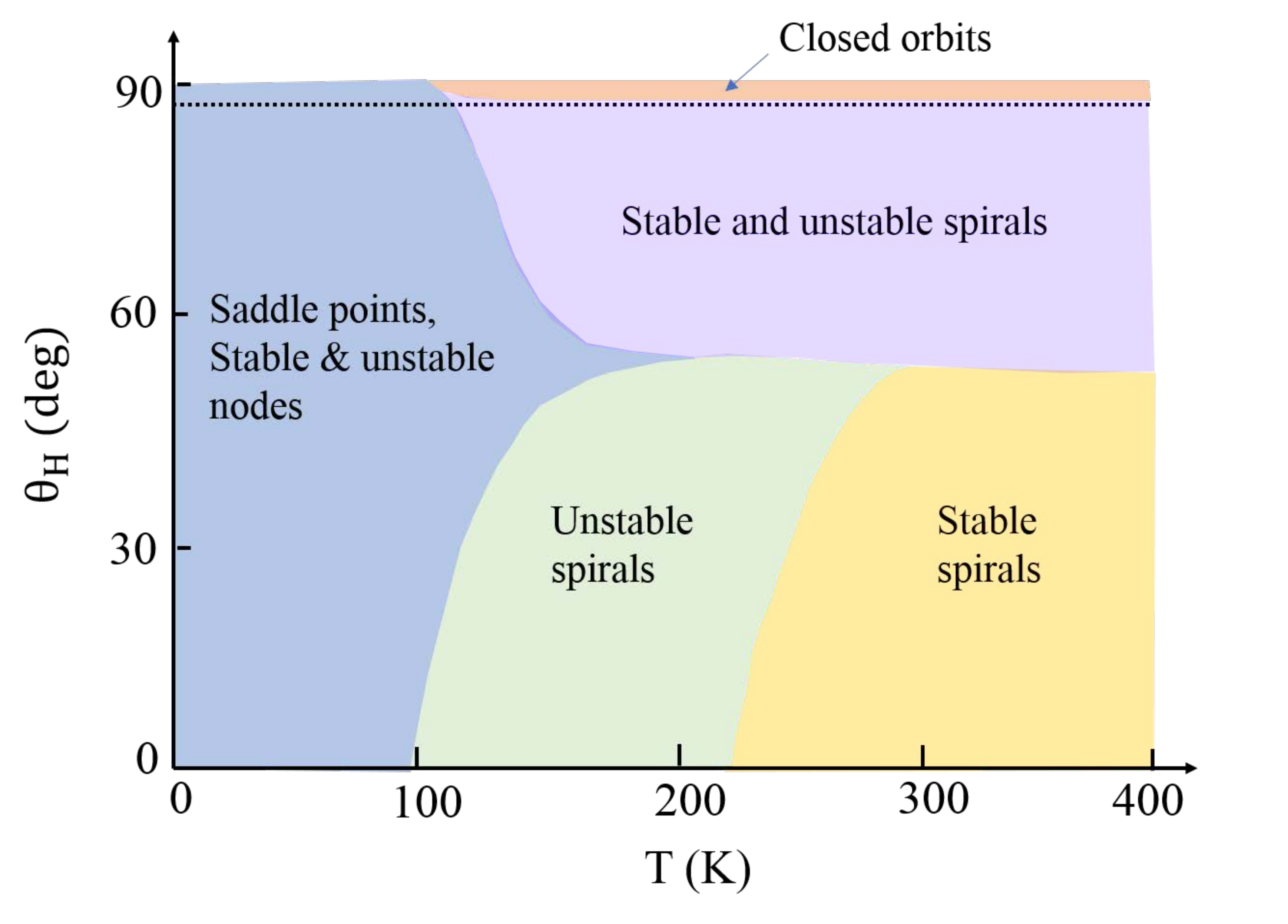}
\caption{Phase diagram of coupled mode dynamics as a function of temperature and field angle of external magnetic field with a fixed magnitude of $15000$~Oe. Closed orbits and several different types of fixed points are shown, such as saddle points, nodes, spirals and etc., and their stability changes across the phase boundaries.  The horizontal black dotted line denotes the field angle of 86$^{\circ}$ (i.e., the zero-crossing point of the nonlinear frequency shift), above which large closed orbits appear at elevated temperatures ($T>100$~K) as the strength of the linear mode coupling grows rapidly.}
\label{fig:phase-diagram}
\end{figure}

Lastly, in Fig.~\ref{fig:phase-diagram}, we show a schematic phase diagram of the dynamic behavior of the coupled mode system as a function of the field angle and temperature, which depicts how changing these control parameters alters the dynamical landscape.

\section{Summary and conclusions}
We theoretically investigated the coupled dynamics of linear spin wave eigenmodes in NC-STOs using a previously derived multi-mode theory~\cite{muduli2012prl,Iacocca2014prb, Zhang16jmmm}. For a simple but experimentally relevant geometry in which the external magnetic field and both equilibrium magnetizations of the free and fixed layers are coplanar, we derived the rate equations that govern the slow dynamics of a subsystem involving several dominant modes, as a generalization of the single mode STO theory proposed earlier by Slavin and Tiberkevich~\cite{slavin2005prl,slavin2008ieeem,slavin2009ieee}. In this particular geometry, we could explicitly calculate the mode-coupling coefficients in the multi-mode theory and transform the system of equations for two dominant modes into an effective two-dimensional driven dynamic system. This allowed us to explore the effect of external field and temperature on the phase portraits of the system and to draw several conclusions about the dynamics of the system. First of all, there is an intimate relation between the non-linear frequency shift and the linear mode coupling [Eq.~(\ref{Eq:R_ij})]. This leads to a minimum in the magnitudes of the mode-coupling coefficients when the non-linear frequency shift is zero, concomitant with a steep change in the phase shift between them. There are profound consequences in the resulting dynamics and phase portraits as the mode-coupling coefficients go through their minimum. The phase portrait of the sub-dynamical system changes rapidly, exhibiting closed orbits, and a set of
different types of fixed points as well as the change  of their stabilities. This is consistent with and explains the observed behavior in STOs~\cite{Dumas2014}. Second, our work explains, through the temperature dependence of the mode-coupling coefficients, how temperature alone can drive the dynamics of the system from one set of orbits and fixed points to another. This is consistent with experimental observations~\cite{Muduli12PRB_T-dep-STO} that to the best of our knowledge have until now eluded explanation; a thermal stochastic field (not included here) only perturbs the orbits about the underlying manifold.  A stochastic field is, however, necessary to induce  mode-hopping~\cite{muduli2012prl,Iacocca2014prb} over saddle points separating orbits. Among other thing, our analysis reveals that increasing the power in the two-mode  subsystem may effectively suppress the linear mode coupling, as the system becomes less perturbed by the interaction between the modes.

Lastly, we stress that the multimode theory that we derived here is based on the expansion of general solutions of modes in terms of the linear combination of eigenmodes, which are propagating spin wave modes in the present case, and hence is not applicable for describing the coupled magnetodynamics involving a localized bullet mode~\cite{slavin2005prl,Iacocca2015prb}.

\begin{acknowledgments}
Work by S. S.-L. Zhang was supported by NSF Grants DMR-1406568 and partly by Department of Energy, Office of Science, Materials Sciences and Engineering Division through Materials Theory Institute. E.I. acknowledges support from the Swedish Research Council, Reg. No. 637-2014-6863. Work by O.H. was supported by the U.S. Department of Energy (DOE), Office of Science, Division of Materials Sciences and Engineering, and O.H. gratefully acknowledges the computing resources provided on Blues and Fusion, high-performance computing clusters operated by the Laboratory Computing Resource Center at Argonne National Laboratory.
\end{acknowledgments}

\appendix
\section{Linear spin wave mode in NC-STO}

In this appendix, we outline the derivation of the profiles of the linear spin wave mode by solving Eq.~(\ref{Eq: EOM-d}) given in the main text.
Separating the time and spatial variables of the wave function profile,
i.e.,
\begin{equation}
d\left( \tilde{r} ,t\right) =C\mathrm{e}^{-\mathrm{i}\omega t}\nu\left( \tilde{r} \right)\,,
\label{Eq: Bessel-u(r)}
\end{equation}%
where $C$ is the normalization coefficient to be determined by boundary conditions, and $\tilde{r}\equiv r/R_c$ is the dimensionless radial distance. Placing Eq.~(\ref{Eq: Bessel-u(r)}) in Eq.~(\ref{Eq: EOM-d}), we arrive at
a zeroth-order Bessel equation for the spatial part of the wave function
\begin{equation}
\tilde{r}^{2}\frac{\mathrm{d}^{2}}{\mathrm{d}\tilde{r}^{2}}\nu\left( \tilde{r}\right) +\tilde{r}\frac{\mathrm{d%
}}{\mathrm{d}\tilde{r}}\nu\left( \tilde{r}\right) +\left( \tilde{\omega}-\tilde{\omega}
_{r}+\mathrm{i}\tilde{\Gamma} _{\alpha }-\mathrm{i}\tilde{\Gamma}
_{J}\right) \tilde{r}^{2}\nu\left( \tilde{r}\right) =0\,,  \label{Eq: Bessel}
\end{equation}%
where $%
\tilde{\omega}_{r}\equiv \omega _{r}/\left( D_{ex}/R_{c}^{2}\right) $, $%
\tilde{\omega}\equiv \omega /\left( D_{ex}/R_{c}^{2}\right) $, $\tilde{\Gamma}
_{\alpha }\equiv \Gamma _{\alpha }/\left( D_{ex}/R_{c}^{2}\right) $
and $\tilde{\Gamma} _{J}=\Gamma _{J}/\left( D_{ex}/R_{c}^{2}\right) $. The
general solution for $\tilde{r}\leq 1$ reads
\begin{equation}
\nu_{<}\left(\tilde{r}\right) =C_{1}J_{0}\left( \kappa _{<}\tilde{r}\right)\,,
\end{equation}%
where $J_{0}$ is the zeroth order Bessel function with
\begin{equation}
\kappa _{<}^{2}=\tilde{\omega} -\tilde{\omega} _{r}+\mathrm{i}\tilde{\Gamma}
_{\alpha }-\mathrm{i}\tilde{\Gamma} _{J}\,.  \label{Eq: kappa<}
\end{equation}%
For $\tilde{r}>1$, we should have an outgoing waves so Hankel function of the first
kind is chosen, i.e.,
\begin{equation}
\nu_{>}\left( \tilde{r}\right) =C_{2}H_{0}^{\left( 1\right) }\left( \kappa _{>}\tilde{r}\right)\,,
\end{equation}%
where
\begin{equation}
\kappa _{>}^{2}=\tilde{\omega}-\tilde{\omega} _{r}+\mathrm{i}\tilde{\Gamma}
_{\alpha }\,.  \label{Eq: kappa>}
\end{equation}%

The coefficients $C_{1}$ and $C_{2}$ are determined by the matching boundary
condition as well as the normalization condition, i.e.,
\begin{equation}
C_{1}J_{0}\left( \kappa _{<}\right) =C_{2}H_{0}^{\left( 1\right)
}\left( \kappa _{>}\right)\,,
\end{equation}%
and%
\begin{equation}
\int_{0}^{R_F/R_c }d\tilde{r}\tilde{r}\left\vert d\left( \tilde{r}\right) \right\vert ^{2}=1\,.
\end{equation}%
Explicitly, we have
\begin{equation}
\left\vert C_{1}\right\vert ^{2}=\frac{\left\vert H_{0}^{\left( 1\right)
}\left( \kappa _{>}\right) \right\vert ^{2}}{\left\vert H_{0}^{\left(
1\right) }\left( \kappa _{>}\right) \right\vert
^{2}\int_{0}^{1}d\tilde{r}\tilde{r}\left\vert d\left( \tilde{r}\right) \right\vert
^{2}+\left\vert J_{0}\left( \kappa _{<}\right) \right\vert
^{2}\int_{1}^{R_F/R_c }d\tilde{r}\tilde{r}\left\vert d\left( \tilde{r}\right) \right\vert ^{2}}\,,
\end{equation}%
and
\begin{equation}
\left\vert C_{1}\right\vert ^{2}=\frac{\left\vert J_{0}\left( \kappa
_{<}\right) \right\vert ^{2}}{\left\vert H_{0}^{\left( 1\right) }\left(
\kappa _{>}\right) \right\vert ^{2}\int_{0}^{1}d\tilde{r}\tilde{r}\left\vert
d\left( \tilde{r}\right) \right\vert ^{2}+\left\vert J_{0}\left( \kappa
_{<}\right) \right\vert ^{2}\int_{1}^{R_F/R_c }d\tilde{r}\tilde{r}\left\vert d\left(
\tilde{r}\right) \right\vert ^{2}}\,.
\end{equation}

By matching the wave functions and their derivatives at $\tilde{r}=1$, we find a
transcendental equation
\begin{equation}
\frac{\kappa _{<}J_{1}\left( \kappa _{<}\right) }{J_{0}\left( \kappa
_{<}\right) }=\frac{\kappa _{>}H_{1}^{(1)}\left( \kappa _{>}\right) }{%
H_{0}^{\left( 1\right) }\left( \kappa _{>}\right) }\,,
\label{Eq: boundary-condition}
\end{equation}%
where we have used the recurrence relation $dZ_{0}\left( x\right)/dx
=-Z_{1}\left( x\right) $ with $Z$ denoting $J$ or $H^{\left( 1\right) }$.
Solving this equation, we can obtain the eigenmodes for a given current
density. There are infinitely many solutions correspond to the excited spin
wave modes with different wave vectors (associated with the number of nodes $%
n$ in the current flowing region). The spin wave frequency in the ultra-thin
limit ($k_{n}d\ll 1$) can be expressed as~\cite{Harte68JAP_demag-thinFilm,Mills99PRB,McMichael98JAP}
\begin{equation}
\omega _{n}=\sqrt{\left( \omega _{H}+D_{ex}k_{n}^{2}\right) \left( \omega
_{H}+D_{ex}k_{n}^{2}+\omega _{M}\cos ^{2}\theta _{M}-\omega _{A}\sin
^{2}\theta _{M}\right) }\,.
\end{equation}%
By solving Eq.~(\ref{Eq: boundary-condition}), we find the wave vectors of
the two lowest spin wave modes are
\begin{equation}
k_{1}=1.76/R_{c}\text{ and }k_{2}=4.61/R_{c}\,.
\end{equation}%
As indicated by Eqs.~(\ref{Eq: kappa<}) and (\ref{Eq: kappa>}), the
wavelengths of excited spin waves in general depend on both damping and
current; in the small damping limit, we recover Slonczewski's result~\cite{slonczewski1999jmmm} of $k_{1}=1.19/R_{c}$ and $k_{2}=4.5/R_{c}$.

\bibliography{STO}

\begin{thebibliography}{30}%
\makeatletter
\providecommand \@ifxundefined [1]{%
 \@ifx{#1\undefined}
}%
\providecommand \@ifnum [1]{%
 \ifnum #1\expandafter \@firstoftwo
 \else \expandafter \@secondoftwo
 \fi
}%
\providecommand \@ifx [1]{%
 \ifx #1\expandafter \@firstoftwo
 \else \expandafter \@secondoftwo
 \fi
}%
\providecommand \natexlab [1]{#1}%
\providecommand \enquote  [1]{``#1''}%
\providecommand \bibnamefont  [1]{#1}%
\providecommand \bibfnamefont [1]{#1}%
\providecommand \citenamefont [1]{#1}%
\providecommand \href@noop [0]{\@secondoftwo}%
\providecommand \href [0]{\begingroup \@sanitize@url \@href}%
\providecommand \@href[1]{\@@startlink{#1}\@@href}%
\providecommand \@@href[1]{\endgroup#1\@@endlink}%
\providecommand \@sanitize@url [0]{\catcode `\\12\catcode `\$12\catcode
  `\&12\catcode `\#12\catcode `\^12\catcode `\_12\catcode `\%12\relax}%
\providecommand \@@startlink[1]{}%
\providecommand \@@endlink[0]{}%
\providecommand \url  [0]{\begingroup\@sanitize@url \@url }%
\providecommand \@url [1]{\endgroup\@href {#1}{\urlprefix }}%
\providecommand \urlprefix  [0]{URL }%
\providecommand \Eprint [0]{\href }%
\@ifxundefined \urlstyle {%
  \providecommand \doi  [0]{\begingroup \@sanitize@url \@doi}%
  \providecommand \@doi [1]{\endgroup \@@startlink {\doibase
  #1}doi:\discretionary {}{}{}#1\@@endlink }%
}{%
  \providecommand \doi  [0]{doi:\discretionary{}{}{}\begingroup
  \urlstyle{rm}\Url }%
}%
\providecommand \doibase [0]{http://dx.doi.org/}%
\providecommand \Doi [0]{\begingroup \@sanitize@url \@Doi }%
\providecommand \@Doi  [1]{\endgroup\@@startlink{\doibase#1}\@@Doi}%
\providecommand \@@Doi [1]{#1\@@endlink}%
\providecommand \selectlanguage [0]{\@gobble}%
\providecommand \bibinfo  [0]{\@secondoftwo}%
\providecommand \bibfield  [0]{\@secondoftwo}%
\providecommand \translation [1]{[#1]}%
\providecommand \BibitemOpen [0]{}%
\providecommand \bibitemStop [0]{}%
\providecommand \bibitemNoStop [0]{.\EOS\space}%
\providecommand \EOS [0]{\spacefactor3000\relax}%
\providecommand \BibitemShut  [1]{\csname bibitem#1\endcsname}%
\bibitem [{\citenamefont {{Slonczewski}}(1996)}]{slonczewski1996jmmm}%
  \BibitemOpen
  \bibfield  {author} {\bibinfo {author} {\bibfnamefont {J.~C.}\ \bibnamefont
  {{Slonczewski}}},\ }\bibfield  {title} {\enquote {\bibinfo {title}
  {{Current-driven excitation of magnetic multilayers}},}\ }\Doi
  {10.1016/0304-8853(96)00062-5} {\bibfield  {journal} {\bibinfo  {journal}
  {\href{http://dx.doi.org/10.1016/0304-8853(96)00062-5}{\emph {J.\ Magn.\
  Magn.\ Mater.}}},\ }\textbf {\bibinfo {volume} {159}},\ \bibinfo {pages}
  {L1}\  (\bibinfo {year} {1996})}\BibitemShut {NoStop}%
\bibitem [{\citenamefont {{Berger}}(1996)}]{berger1996prb}%
  \BibitemOpen
  \bibfield  {author} {\bibinfo {author} {\bibfnamefont {L.}~\bibnamefont
  {{Berger}}},\ }\bibfield  {title} {\enquote {\bibinfo {title} {{Emission of
  spin waves by a magnetic multilayer traversed by a current}},}\ }\Doi
  {10.1103/PhysRevB.54.9353} {\bibfield  {journal} {\bibinfo  {journal}
  {\href{http://dx.doi.org/10.1103/PhysRevB.54.9353}{\emph {\prb}}},\ }\textbf
  {\bibinfo {volume} {54}},\ \bibinfo {pages} {9353}\  (\bibinfo {year}
  {1996})}\BibitemShut {NoStop}%
\bibitem [{\citenamefont {{Silva}}\ and\ \citenamefont
  {{Rippard}}(2008)}]{silva2008jmmm}%
  \BibitemOpen
  \bibfield  {author} {\bibinfo {author} {\bibfnamefont {T.~J.}\ \bibnamefont
  {{Silva}}}\ and\ \bibinfo {author} {\bibfnamefont {W.~H.}\ \bibnamefont
  {{Rippard}}},\ }\bibfield  {title} {\enquote {\bibinfo {title} {{Developments
  in nano-oscillators based upon spin-transfer point-contact devices}},}\ }\Doi
  {10.1016/j.jmmm.2007.12.022} {\bibfield  {journal} {\bibinfo  {journal}
  {\href{http://dx.doi.org/10.1016/j.jmmm.2007.12.022}{\emph {J.\ Magn.\ Magn.\
  Mater.}}},\ }\textbf {\bibinfo {volume} {320}},\ \bibinfo {pages} {1260}\
  (\bibinfo {year} {2008})}\BibitemShut {NoStop}%
\bibitem [{\citenamefont {Dumas}\ \emph {et~al.}(2014)\citenamefont {Dumas},
  \citenamefont {Sani}, \citenamefont {Mohseni}, \citenamefont {Iacocca},
  \citenamefont {Pogoryelov}, \citenamefont {Muduli}, \citenamefont {Chung},
  \citenamefont {D\"{u}rrenfeld},\ and\ \citenamefont
  {\AA{}kerman}}]{Dumas2014}%
  \BibitemOpen
  \bibfield  {author} {\bibinfo {author} {\bibfnamefont {R.~K.}\ \bibnamefont
  {Dumas}}, \bibinfo {author} {\bibfnamefont {S.~R.}\ \bibnamefont {Sani}},
  \bibinfo {author} {\bibfnamefont {S.~M.}\ \bibnamefont {Mohseni}}, \bibinfo
  {author} {\bibfnamefont {E.}~\bibnamefont {Iacocca}}, \bibinfo {author}
  {\bibfnamefont {Y.}~\bibnamefont {Pogoryelov}}, \bibinfo {author}
  {\bibfnamefont {P.~K.}\ \bibnamefont {Muduli}}, \bibinfo {author}
  {\bibfnamefont {S.}~\bibnamefont {Chung}}, \bibinfo {author} {\bibfnamefont
  {P.}~\bibnamefont {D\"{u}rrenfeld}}, \ and\ \bibinfo {author} {\bibfnamefont
  {J.}~\bibnamefont {\AA{}kerman}},\ }\bibfield  {title} {\enquote {\bibinfo
  {title} {Recent advances in nanocontact spin-torque oscillators},}\ }\Doi
  {10.1109/TMAG.2014.2305762} {\bibfield  {journal} {\bibinfo  {journal}
  {\href{http://dx.doi.org/10.1109/TMAG.2014.2305762}{\emph {Magnetics, IEEE
  Transactions on}}},\ }\textbf {\bibinfo {volume} {50}},\ \bibinfo {pages}
  {1}\  (\bibinfo {year} {2014})}\BibitemShut {NoStop}%
\bibitem [{\citenamefont {Slavin}\ and\ \citenamefont
  {Tiberkevich}(2009)}]{slavin2009ieee}%
  \BibitemOpen
  \bibfield  {author} {\bibinfo {author} {\bibfnamefont {A.}~\bibnamefont
  {Slavin}}\ and\ \bibinfo {author} {\bibfnamefont {V.}~\bibnamefont
  {Tiberkevich}},\ }\bibfield  {title} {\enquote {\bibinfo {title} {Nonlinear
  auto-oscillator theory of microwave generation by spin-polarized current},}\
  }\href@noop {} {\bibfield  {journal} {\bibinfo  {journal} {\emph {Magnetics,
  IEEE Transactions on}},\ }\textbf {\bibinfo {volume} {45}},\ \bibinfo {pages}
  {1875 }\  (\bibinfo {year} {2009})}\BibitemShut {NoStop}%
\bibitem [{\citenamefont {Bertotti}\ \emph {et~al.}(2005)\citenamefont
  {Bertotti}, \citenamefont {Serpico}, \citenamefont {Mayergoyz}, \citenamefont
  {Magni}, \citenamefont {d'Aquino},\ and\ \citenamefont
  {Bonin}}]{Bertotti2005}%
  \BibitemOpen
  \bibfield  {author} {\bibinfo {author} {\bibfnamefont {G.}~\bibnamefont
  {Bertotti}}, \bibinfo {author} {\bibfnamefont {C.}~\bibnamefont {Serpico}},
  \bibinfo {author} {\bibfnamefont {I.~D.}\ \bibnamefont {Mayergoyz}}, \bibinfo
  {author} {\bibfnamefont {A.}~\bibnamefont {Magni}}, \bibinfo {author}
  {\bibfnamefont {M.}~\bibnamefont {d'Aquino}}, \ and\ \bibinfo {author}
  {\bibfnamefont {R.}~\bibnamefont {Bonin}},\ }\bibfield  {title} {\enquote
  {\bibinfo {title} {Magnetization switching and microwave oscillations in
  nanomagnets driven by spin-polarized currents},}\ }\href@noop {} {\bibfield
  {journal} {\bibinfo  {journal} {\emph {Phys. Rev. Lett.}},\ }\textbf
  {\bibinfo {volume} {94}},\ \bibinfo {pages} {127206}\  (\bibinfo {year}
  {2005})}\BibitemShut {NoStop}%
\bibitem [{\citenamefont {{de Aguiar}}\ \emph {et~al.}(2007)\citenamefont {{de
  Aguiar}}, \citenamefont {{Azevedo}},\ and\ \citenamefont
  {{Rezende}}}]{deAguiar2007prb}%
  \BibitemOpen
  \bibfield  {author} {\bibinfo {author} {\bibfnamefont {F.~M.}\ \bibnamefont
  {{de Aguiar}}}, \bibinfo {author} {\bibfnamefont {A.}~\bibnamefont
  {{Azevedo}}}, \ and\ \bibinfo {author} {\bibfnamefont {S.~M.}\ \bibnamefont
  {{Rezende}}},\ }\bibfield  {title} {\enquote {\bibinfo {title} {{Theory of a
  two-mode spin torque nanooscillator}},}\ }\Doi {10.1103/PhysRevB.75.132404}
  {\bibfield  {journal} {\bibinfo  {journal}
  {\href{http://dx.doi.org/10.1103/PhysRevB.75.132404}{\emph {Phys.\ Rev.\
  B}}},\ }\textbf {\bibinfo {volume} {75}},\ \bibinfo {pages} {132404}\
  (\bibinfo {year} {2007})}\BibitemShut {NoStop}%
\bibitem [{\citenamefont {{Krivorotov}}\ \emph {et~al.}(2008)\citenamefont
  {{Krivorotov}}, \citenamefont {{Emley}}, \citenamefont {{Buhrman}},\ and\
  \citenamefont {{Ralph}}}]{krivorotov2008prb}%
  \BibitemOpen
  \bibfield  {author} {\bibinfo {author} {\bibfnamefont {I.~N.}\ \bibnamefont
  {{Krivorotov}}}, \bibinfo {author} {\bibfnamefont {N.~C.}\ \bibnamefont
  {{Emley}}}, \bibinfo {author} {\bibfnamefont {R.~A.}\ \bibnamefont
  {{Buhrman}}}, \ and\ \bibinfo {author} {\bibfnamefont {D.~C.}\ \bibnamefont
  {{Ralph}}},\ }\bibfield  {title} {\enquote {\bibinfo {title} {{Time-domain
  studies of very-large-angle magnetization dynamics excited by spin transfer
  torques}},}\ }\Doi {10.1103/PhysRevB.77.054440} {\bibfield  {journal}
  {\bibinfo  {journal}
  {\href{http://dx.doi.org/10.1103/PhysRevB.77.054440}{\emph {Phys.\ Rev.\
  B}}},\ }\textbf {\bibinfo {volume} {77}},\ \bibinfo {pages} {054440}\
  (\bibinfo {year} {2008})}\BibitemShut {NoStop}%
\bibitem [{\citenamefont {{Muduli}}\ \emph {et~al.}(2012)\citenamefont
  {{Muduli}}, \citenamefont {{Heinonen}},\ and\ \citenamefont
  {{{\AA}kerman}}}]{muduli2012prl}%
  \BibitemOpen
  \bibfield  {author} {\bibinfo {author} {\bibfnamefont {P.~K.}\ \bibnamefont
  {{Muduli}}}, \bibinfo {author} {\bibfnamefont {O.~G.}\ \bibnamefont
  {{Heinonen}}}, \ and\ \bibinfo {author} {\bibfnamefont {J.}~\bibnamefont
  {{{\AA}kerman}}},\ }\bibfield  {title} {\enquote {\bibinfo {title}
  {{Decoherence and Mode Hopping in a Magnetic Tunnel Junction Based Spin
  Torque Oscillator}},}\ }\Doi {10.1103/PhysRevLett.108.207203} {\bibfield
  {journal} {\bibinfo  {journal}
  {\href{http://dx.doi.org/10.1103/PhysRevLett.108.207203}{\emph {Phys.\ Rev.\
  Lett.}}},\ }\textbf {\bibinfo {volume} {108}},\ \bibinfo {eid} {207203}\
  (\bibinfo {year} {2012})}\BibitemShut {NoStop}%
\bibitem [{\citenamefont {Muduli}\ \emph {et~al.}(2012)\citenamefont {Muduli},
  \citenamefont {Heinonen},\ and\ \citenamefont
  {\AA{}kerman}}]{Muduli12PRB_T-dep-STO}%
  \BibitemOpen
  \bibfield  {author} {\bibinfo {author} {\bibfnamefont {P.~K.}\ \bibnamefont
  {Muduli}}, \bibinfo {author} {\bibfnamefont {O.~G.}\ \bibnamefont
  {Heinonen}}, \ and\ \bibinfo {author} {\bibfnamefont {J.}~\bibnamefont
  {\AA{}kerman}},\ }\bibfield  {title} {\enquote {\bibinfo {title} {Temperature
  dependence of linewidth in nanocontact based spin torque oscillators: Effect
  of multiple oscillatory modes},}\ }\Doi {10.1103/PhysRevB.86.174408}
  {\bibfield  {journal} {\bibinfo  {journal}
  {\href{http://dx.doi.org/10.1103/PhysRevB.86.174408}{\emph {Phys. Rev. B}}},\
  }\textbf {\bibinfo {volume} {86}},\ \bibinfo {pages} {174408}\  (\bibinfo
  {year} {2012})}\BibitemShut {NoStop}%
\bibitem [{\citenamefont {{Bonetti}}\ \emph {et~al.}(2010)\citenamefont
  {{Bonetti}}, \citenamefont {{Tiberkevich}}, \citenamefont {{Consolo}},
  \citenamefont {{Finocchio}}, \citenamefont {{Muduli}}, \citenamefont
  {{Mancoff}}, \citenamefont {{Slavin}},\ and\ \citenamefont
  {{{\AA}kerman}}}]{bonetti2010prl}%
  \BibitemOpen
  \bibfield  {author} {\bibinfo {author} {\bibfnamefont {S.}~\bibnamefont
  {{Bonetti}}}, \bibinfo {author} {\bibfnamefont {V.}~\bibnamefont
  {{Tiberkevich}}}, \bibinfo {author} {\bibfnamefont {G.}~\bibnamefont
  {{Consolo}}}, \bibinfo {author} {\bibfnamefont {G.}~\bibnamefont
  {{Finocchio}}}, \bibinfo {author} {\bibfnamefont {P.}~\bibnamefont
  {{Muduli}}}, \bibinfo {author} {\bibfnamefont {F.}~\bibnamefont {{Mancoff}}},
  \bibinfo {author} {\bibfnamefont {A.}~\bibnamefont {{Slavin}}}, \ and\
  \bibinfo {author} {\bibfnamefont {J.}~\bibnamefont {{{\AA}kerman}}},\
  }\bibfield  {title} {\enquote {\bibinfo {title} {{Experimental Evidence of
  Self-Localized and Propagating Spin Wave Modes in Obliquely Magnetized
  Current-Driven Nanocontacts}},}\ }\Doi {10.1103/PhysRevLett.105.217204}
  {\bibfield  {journal} {\bibinfo  {journal}
  {\href{http://dx.doi.org/10.1103/PhysRevLett.105.217204}{\emph {Phys.\ Rev.\
  Lett.}}},\ }\textbf {\bibinfo {volume} {105}},\ \bibinfo {pages} {217204}\
  (\bibinfo {year} {2010})}\BibitemShut {NoStop}%
\bibitem [{\citenamefont {{Bonetti}}\ \emph {et~al.}(2012)\citenamefont
  {{Bonetti}}, \citenamefont {{Puliafito}}, \citenamefont {{Consolo}},
  \citenamefont {{Tiberkevich}}, \citenamefont {{Slavin}},\ and\ \citenamefont
  {{{\AA}kerman}}}]{bonetti2012prb}%
  \BibitemOpen
  \bibfield  {author} {\bibinfo {author} {\bibfnamefont {S.}~\bibnamefont
  {{Bonetti}}}, \bibinfo {author} {\bibfnamefont {V.}~\bibnamefont
  {{Puliafito}}}, \bibinfo {author} {\bibfnamefont {G.}~\bibnamefont
  {{Consolo}}}, \bibinfo {author} {\bibfnamefont {V.~S.}\ \bibnamefont
  {{Tiberkevich}}}, \bibinfo {author} {\bibfnamefont {A.~N.}\ \bibnamefont
  {{Slavin}}}, \ and\ \bibinfo {author} {\bibfnamefont {J.}~\bibnamefont
  {{{\AA}kerman}}},\ }\bibfield  {title} {\enquote {\bibinfo {title} {{Power
  and linewidth of propagating and localized modes in nanocontact spin-torque
  oscillators}},}\ }\Doi {10.1103/PhysRevB.85.174427} {\bibfield  {journal}
  {\bibinfo  {journal}
  {\href{http://dx.doi.org/10.1103/PhysRevB.85.174427}{\emph {Phys.\ Rev.\
  B}}},\ }\textbf {\bibinfo {volume} {85}},\ \bibinfo {eid} {174427}\
  (\bibinfo {year} {2012})}\BibitemShut {NoStop}%
\bibitem [{\citenamefont {Dumas}\ \emph {et~al.}(2013)\citenamefont {Dumas},
  \citenamefont {Iacocca}, \citenamefont {Bonetti}, \citenamefont {Sani},
  \citenamefont {Mohseni}, \citenamefont {Eklund}, \citenamefont {Persson},
  \citenamefont {Heinonen},\ and\ \citenamefont {{\AA}kerman}}]{dumas2013prl}%
  \BibitemOpen
  \bibfield  {author} {\bibinfo {author} {\bibfnamefont {R.}~\bibnamefont
  {Dumas}}, \bibinfo {author} {\bibfnamefont {E.}~\bibnamefont {Iacocca}},
  \bibinfo {author} {\bibfnamefont {S.}~\bibnamefont {Bonetti}}, \bibinfo
  {author} {\bibfnamefont {S.}~\bibnamefont {Sani}}, \bibinfo {author}
  {\bibfnamefont {S.}~\bibnamefont {Mohseni}}, \bibinfo {author} {\bibfnamefont
  {A.}~\bibnamefont {Eklund}}, \bibinfo {author} {\bibfnamefont
  {J.}~\bibnamefont {Persson}}, \bibinfo {author} {\bibfnamefont
  {O.}~\bibnamefont {Heinonen}}, \ and\ \bibinfo {author} {\bibfnamefont
  {J.}~\bibnamefont {{\AA}kerman}},\ }\href@noop {} {\bibfield  {journal}
  {\bibinfo  {journal} {\emph {Phys.\ Rev.\ Lett.}},\ }\textbf {\bibinfo
  {volume} {110}},\ \bibinfo {pages} {257202}\  (\bibinfo {year}
  {2013})}\BibitemShut {NoStop}%
\bibitem [{\citenamefont {Iacocca}\ \emph {et~al.}(2015)\citenamefont
  {Iacocca}, \citenamefont {D\"urrenfeld}, \citenamefont {Heinonen},
  \citenamefont {\AA{}kerman},\ and\ \citenamefont {Dumas}}]{Iacocca2015prb}%
  \BibitemOpen
  \bibfield  {author} {\bibinfo {author} {\bibfnamefont {E.}~\bibnamefont
  {Iacocca}}, \bibinfo {author} {\bibfnamefont {P.}~\bibnamefont
  {D\"urrenfeld}}, \bibinfo {author} {\bibfnamefont {O.}~\bibnamefont
  {Heinonen}}, \bibinfo {author} {\bibfnamefont {J.}~\bibnamefont
  {\AA{}kerman}}, \ and\ \bibinfo {author} {\bibfnamefont {R.~K.}\ \bibnamefont
  {Dumas}},\ }\bibfield  {title} {\enquote {\bibinfo {title} {Mode-coupling
  mechanisms in nanocontact spin-torque oscillators},}\ }\Doi
  {10.1103/PhysRevB.91.104405} {\bibfield  {journal} {\bibinfo  {journal}
  {\href{http://dx.doi.org/10.1103/PhysRevB.91.104405}{\emph {Phys. Rev. B}}},\
  }\textbf {\bibinfo {volume} {91}},\ \bibinfo {pages} {104405}\  (\bibinfo
  {year} {2015})}\BibitemShut {NoStop}%
\bibitem [{\citenamefont {Beri}\ \emph {et~al.}(2008)\citenamefont {Beri},
  \citenamefont {Gelens}, \citenamefont {Mestre}, \citenamefont {{van der
  Sande}}, \citenamefont {{Verschaffelt}}, \citenamefont {{Scir\`e}},
  \citenamefont {Mezosi}, \citenamefont {Sorel},\ and\ \citenamefont
  {Danckaert}}]{beri2008prl}%
  \BibitemOpen
  \bibfield  {author} {\bibinfo {author} {\bibfnamefont {S.}~\bibnamefont
  {Beri}}, \bibinfo {author} {\bibfnamefont {L.}~\bibnamefont {Gelens}},
  \bibinfo {author} {\bibfnamefont {M.}~\bibnamefont {Mestre}}, \bibinfo
  {author} {\bibfnamefont {G.}~\bibnamefont {{van der Sande}}}, \bibinfo
  {author} {\bibfnamefont {G.}~\bibnamefont {{Verschaffelt}}}, \bibinfo
  {author} {\bibfnamefont {G.}~\bibnamefont {{Scir\`e}}}, \bibinfo {author}
  {\bibfnamefont {G.}~\bibnamefont {Mezosi}}, \bibinfo {author} {\bibfnamefont
  {M.}~\bibnamefont {Sorel}}, \ and\ \bibinfo {author} {\bibfnamefont
  {J.}~\bibnamefont {Danckaert}},\ }\href@noop {} {\bibfield  {journal}
  {\bibinfo  {journal} {\emph {Phys.\ Rev.\ Lett.}},\ }\textbf {\bibinfo
  {volume} {101}},\ \bibinfo {pages} {093903}\  (\bibinfo {year}
  {2008})}\BibitemShut {NoStop}%
\bibitem [{\citenamefont {van~der Sande}\ \emph {et~al.}(2008)\citenamefont
  {van~der Sande}, \citenamefont {Gelens}, \citenamefont {Tassin},\ and\
  \citenamefont {Scir\`e}}]{vanderSande}%
  \BibitemOpen
  \bibfield  {author} {\bibinfo {author} {\bibfnamefont {G.}~\bibnamefont
  {van~der Sande}}, \bibinfo {author} {\bibfnamefont {L.}~\bibnamefont
  {Gelens}}, \bibinfo {author} {\bibfnamefont {P.}~\bibnamefont {Tassin}}, \
  and\ \bibinfo {author} {\bibfnamefont {J.}~\bibnamefont {Scir\`e},
  \bibfnamefont {A.~aand~Danckaert}},\ }\href@noop {} {\bibfield  {journal}
  {\bibinfo  {journal} {\emph {J. Phys. B: At. Mol. Opt. Phys.}},\ }\textbf
  {\bibinfo {volume} {41}},\ \bibinfo {pages} {095402}\  (\bibinfo {year}
  {2008})}\BibitemShut {NoStop}%
\bibitem [{\citenamefont {Iacocca}\ \emph {et~al.}(2014)\citenamefont
  {Iacocca}, \citenamefont {Heinonen}, \citenamefont {Muduli},\ and\
  \citenamefont {\AA{}kerman}}]{Iacocca2014prb}%
  \BibitemOpen
  \bibfield  {author} {\bibinfo {author} {\bibfnamefont {E.}~\bibnamefont
  {Iacocca}}, \bibinfo {author} {\bibfnamefont {O.}~\bibnamefont {Heinonen}},
  \bibinfo {author} {\bibfnamefont {P.~K.}\ \bibnamefont {Muduli}}, \ and\
  \bibinfo {author} {\bibfnamefont {J.}~\bibnamefont {\AA{}kerman}},\
  }\bibfield  {title} {\enquote {\bibinfo {title} {Generation linewidth of
  mode-hopping spin torque oscillators},}\ }\Doi {10.1103/PhysRevB.89.054402}
  {\bibfield  {journal} {\bibinfo  {journal}
  {\href{http://dx.doi.org/10.1103/PhysRevB.89.054402}{\emph {Phys. Rev. B}}},\
  }\textbf {\bibinfo {volume} {89}},\ \bibinfo {pages} {054402}\  (\bibinfo
  {year} {2014})}\BibitemShut {NoStop}%
\bibitem [{\citenamefont {Sharma}\ \emph {et~al.}(2014)\citenamefont {Sharma},
  \citenamefont {D{\"u}rrenfeld}, \citenamefont {Iacocca}, \citenamefont
  {Heinonen},\ and\ \citenamefont {{\AA}kerman}}]{Sharma2014APL}%
  \BibitemOpen
  \bibfield  {author} {\bibinfo {author} {\bibfnamefont {N.}~\bibnamefont
  {Sharma}}, \bibinfo {author} {\bibfnamefont {P.}~\bibnamefont
  {D{\"u}rrenfeld}}, \bibinfo {author} {\bibfnamefont {E.}~\bibnamefont
  {Iacocca}}, \bibinfo {author} {\bibfnamefont {O.}~\bibnamefont {Heinonen}}, \
  and\ \bibinfo {author} {\bibfnamefont {J.}~\bibnamefont {{\AA}kerman}},\
  }\href@noop {} {\bibfield  {journal} {\bibinfo  {journal} {\emph {Appl. Phys.
  Lett.}},\ }\textbf {\bibinfo {volume} {105}},\ \bibinfo {pages} {132404}\
  (\bibinfo {year} {2014})}\BibitemShut {NoStop}%
\bibitem [{\citenamefont {Eklund}\ \emph {et~al.}(2014)\citenamefont {Eklund},
  \citenamefont {Bonetti}, \citenamefont {Sani}, \citenamefont {Majid~Mohseni},
  \citenamefont {Persson}, \citenamefont {Chung}, \citenamefont {Amir
  Hossein~Banuazizi}, \citenamefont {Iacocca}, \citenamefont {\"{O}stling},
  \citenamefont {\AA{}kerman},\ and\ \citenamefont {Gunnar~Malm}}]{Eklund2014}%
  \BibitemOpen
  \bibfield  {author} {\bibinfo {author} {\bibfnamefont {A.}~\bibnamefont
  {Eklund}}, \bibinfo {author} {\bibfnamefont {S.}~\bibnamefont {Bonetti}},
  \bibinfo {author} {\bibfnamefont {S.~R.}\ \bibnamefont {Sani}}, \bibinfo
  {author} {\bibfnamefont {S.}~\bibnamefont {Majid~Mohseni}}, \bibinfo {author}
  {\bibfnamefont {J.}~\bibnamefont {Persson}}, \bibinfo {author} {\bibfnamefont
  {S.}~\bibnamefont {Chung}}, \bibinfo {author} {\bibfnamefont
  {S.}~\bibnamefont {Amir Hossein~Banuazizi}}, \bibinfo {author} {\bibfnamefont
  {E.}~\bibnamefont {Iacocca}}, \bibinfo {author} {\bibfnamefont
  {M.}~\bibnamefont {\"{O}stling}}, \bibinfo {author} {\bibfnamefont
  {J.}~\bibnamefont {\AA{}kerman}}, \ and\ \bibinfo {author} {\bibfnamefont
  {B.}~\bibnamefont {Gunnar~Malm}},\ }\bibfield  {title} {\enquote {\bibinfo
  {title} {Dependence of the colored frequency noise in spin torque oscillators
  on current and magnetic field},}\ }\Doi {http://dx.doi.org/10.1063/1.4867257}
  {\bibfield  {journal} {\bibinfo  {journal}
  {\href{http://dx.doi.org/http://dx.doi.org/10.1063/1.4867257}{\emph {Applied
  Physics Letters}}},\ }\textbf {\bibinfo {volume} {104}},\ \bibinfo {eid}
  {092405}\  (\bibinfo {year} {2014})}\BibitemShut {NoStop}%
\bibitem [{\citenamefont {Zhang}\ \emph {et~al.}(2016)\citenamefont {Zhang},
  \citenamefont {Zhou}, \citenamefont {Li},\ and\ \citenamefont
  {Heinonen}}]{Zhang16jmmm}%
  \BibitemOpen
  \bibfield  {author} {\bibinfo {author} {\bibfnamefont {S.~S.-L.}\
  \bibnamefont {Zhang}}, \bibinfo {author} {\bibfnamefont {Y.}~\bibnamefont
  {Zhou}}, \bibinfo {author} {\bibfnamefont {D.}~\bibnamefont {Li}}, \ and\
  \bibinfo {author} {\bibfnamefont {O.}~\bibnamefont {Heinonen}},\ }\bibfield
  {title} {\enquote {\bibinfo {title} {Mode coupling in spin torque
  oscillators},}\ }\href@noop {} {\bibfield  {journal} {\bibinfo  {journal}
  {\emph {J. Magn. Magn. Mater.}},\ }\textbf {\bibinfo {volume} {414}},\
  \bibinfo {pages} {227 }\  (\bibinfo {year} {2016})}\BibitemShut {NoStop}%
\bibitem [{\citenamefont {{Slonczewski}}(1999)}]{slonczewski1999jmmm}%
  \BibitemOpen
  \bibfield  {author} {\bibinfo {author} {\bibfnamefont {J.~C.}\ \bibnamefont
  {{Slonczewski}}},\ }\bibfield  {title} {\enquote {\bibinfo {title}
  {{Excitation of spin waves by an electric current}},}\ }\Doi
  {10.1016/S0304-8853(99)00043-8} {\bibfield  {journal} {\bibinfo  {journal}
  {\href{http://dx.doi.org/10.1016/S0304-8853(99)00043-8}{\emph {J.\ Magn.\
  Magn.\ Mater.}}},\ }\textbf {\bibinfo {volume} {195}},\ \bibinfo {pages}
  {261}\  (\bibinfo {year} {1999})}\BibitemShut {NoStop}%
\bibitem [{Note1()}]{Note1}%
  \BibitemOpen
  \bibinfo {note} {We note that in experiments, the application of a large
  external field will slightly tilt the magnetization of the fixed layer out of
  plane. We will ignore such out-of-plane components as they will be small and
  will primarily only lead to a small renormalization of the STT. Similarly, we
  are ignoring field-like STT as it would slightly renormalize the external
  magnetic field.}\BibitemShut {Stop}%
\bibitem [{\citenamefont {Slavin}\ and\ \citenamefont
  {Tiberkevich}(2008)}]{slavin2008ieeem}%
  \BibitemOpen
  \bibfield  {author} {\bibinfo {author} {\bibfnamefont {A.}~\bibnamefont
  {Slavin}}\ and\ \bibinfo {author} {\bibfnamefont {V.}~\bibnamefont
  {Tiberkevich}},\ }\bibfield  {title} {\enquote {\bibinfo {title} {Excitation
  of spin waves by spin-polarized current in magnetic nano-structures},}\ }\Doi
  {10.1109/TMAG.2008.924537} {\bibfield  {journal} {\bibinfo  {journal}
  {\href{http://dx.doi.org/10.1109/TMAG.2008.924537}{\emph {IEEE Trans.
  Magn.}}},\ }\textbf {\bibinfo {volume} {44}},\ \bibinfo {pages} {1916}\
  (\bibinfo {year} {2008})}\BibitemShut {NoStop}%
\bibitem [{\citenamefont {Krasitskii}(1990)}]{Krasitskii1990JETP}%
  \BibitemOpen
  \bibfield  {author} {\bibinfo {author} {\bibfnamefont {V.~P.}\ \bibnamefont
  {Krasitskii}},\ }\bibfield  {title} {\enquote {\bibinfo {title} {Canonical
  transformation in a theory of weakly nonlinear waves with a nondecay
  dispersion law},}\ }\href@noop {} {\bibfield  {journal} {\bibinfo  {journal}
  {\emph {Sov. Phys. JETP}},\ }\textbf {\bibinfo {volume} {71}},\ \bibinfo
  {pages} {921}\  (\bibinfo {year} {1990})}\BibitemShut {NoStop}%
\bibitem [{\citenamefont {{L'vov}}(1994)}]{Lvov}%
  \BibitemOpen
  \bibfield  {author} {\bibinfo {author} {\bibfnamefont {V.~S.}\ \bibnamefont
  {{L'vov}}},\ }\href@noop {} {\emph {\bibinfo {title} {Wave Turbulence Under
  Parametric Excitations: Applications to Magnets}}}\ (\bibinfo  {publisher}
  {Springer},\ \bibinfo {address} {Berlin},\ \bibinfo {year}
  {1994})\BibitemShut {NoStop}%
\bibitem [{\citenamefont {{Slavin}}\ and\ \citenamefont
  {{Tiberkevich}}(2005)}]{slavin2005prl}%
  \BibitemOpen
  \bibfield  {author} {\bibinfo {author} {\bibfnamefont {A.}~\bibnamefont
  {{Slavin}}}\ and\ \bibinfo {author} {\bibfnamefont {V.}~\bibnamefont
  {{Tiberkevich}}},\ }\bibfield  {title} {\enquote {\bibinfo {title} {{Spin
  Wave Mode Excited by Spin-Polarized Current in a Magnetic Nanocontact is a
  Standing Self-Localized Wave Bullet}},}\ }\Doi
  {10.1103/PhysRevLett.95.237201} {\bibfield  {journal} {\bibinfo  {journal}
  {\href{http://dx.doi.org/10.1103/PhysRevLett.95.237201}{\emph {Phys.\ Rev.\
  Lett.}}},\ }\textbf {\bibinfo {volume} {95}},\ \bibinfo {pages} {237201}\
  (\bibinfo {year} {2005})}\BibitemShut {NoStop}%
\bibitem [{\citenamefont {Krivosik}\ and\ \citenamefont
  {Patton}(2010)}]{Krivosik2010}%
  \BibitemOpen
  \bibfield  {author} {\bibinfo {author} {\bibfnamefont {P.}~\bibnamefont
  {Krivosik}}\ and\ \bibinfo {author} {\bibfnamefont {C.~E.}\ \bibnamefont
  {Patton}},\ }\bibfield  {title} {\enquote {\bibinfo {title} {Hamiltonian
  formulation of nonlinear spin-wave dynamics: Theory and applications},}\
  }\Doi {10.1103/PhysRevB.82.184428} {\bibfield  {journal} {\bibinfo  {journal}
  {\href{http://dx.doi.org/10.1103/PhysRevB.82.184428}{\emph {Phys. Rev. B}}},\
  }\textbf {\bibinfo {volume} {82}},\ \bibinfo {pages} {184428}\  (\bibinfo
  {year} {2010})}\BibitemShut {NoStop}%
\bibitem [{\citenamefont {Harte}(1968)}]{Harte68JAP_demag-thinFilm}%
  \BibitemOpen
  \bibfield  {author} {\bibinfo {author} {\bibfnamefont {K.~J.}\ \bibnamefont
  {Harte}},\ }\bibfield  {title} {\enquote {\bibinfo {title} {Theory of
  magnetization ripple in ferromagnetic films},}\ }\Doi
  {http://dx.doi.org/10.1063/1.1656388} {\bibfield  {journal} {\bibinfo
  {journal} {\href{http://dx.doi.org/http://dx.doi.org/10.1063/1.1656388}{\emph
  {J. Appl. Phys.}}},\ }\textbf {\bibinfo {volume} {39}},\ \bibinfo {pages}
  {1503}\  (\bibinfo {year} {1968})}\BibitemShut {NoStop}%
\bibitem [{\citenamefont {Arias}\ and\ \citenamefont
  {Mills}(1999)}]{Mills99PRB}%
  \BibitemOpen
  \bibfield  {author} {\bibinfo {author} {\bibfnamefont {R.}~\bibnamefont
  {Arias}}\ and\ \bibinfo {author} {\bibfnamefont {D.~L.}\ \bibnamefont
  {Mills}},\ }\bibfield  {title} {\enquote {\bibinfo {title} {Extrinsic
  contributions to the ferromagnetic resonance response of ultrathin films},}\
  }\Doi {10.1103/PhysRevB.60.7395} {\bibfield  {journal} {\bibinfo  {journal}
  {\href{http://dx.doi.org/10.1103/PhysRevB.60.7395}{\emph {Phys. Rev. B}}},\
  }\textbf {\bibinfo {volume} {60}},\ \bibinfo {pages} {7395}\  (\bibinfo
  {year} {1999})}\BibitemShut {NoStop}%
\bibitem [{\citenamefont {McMichael}\ \emph {et~al.}(1998)\citenamefont
  {McMichael}, \citenamefont {Stiles}, \citenamefont {Chen},\ and\
  \citenamefont {Egelhoff}}]{McMichael98JAP}%
  \BibitemOpen
  \bibfield  {author} {\bibinfo {author} {\bibfnamefont {R.~D.}\ \bibnamefont
  {McMichael}}, \bibinfo {author} {\bibfnamefont {M.~D.}\ \bibnamefont
  {Stiles}}, \bibinfo {author} {\bibfnamefont {P.~J.}\ \bibnamefont {Chen}}, \
  and\ \bibinfo {author} {\bibfnamefont {W.~F.}\ \bibnamefont {Egelhoff}},\
  }\bibfield  {title} {\enquote {\bibinfo {title} {Ferromagnetic resonance
  linewidth in thin films coupled to nio},}\ }\Doi
  {http://dx.doi.org/10.1063/1.367725} {\bibfield  {journal} {\bibinfo
  {journal} {\href{http://dx.doi.org/http://dx.doi.org/10.1063/1.367725}{\emph
  {Journal of Applied Physics}}},\ }\textbf {\bibinfo {volume} {83}},\ \bibinfo
  {pages} {7037}\  (\bibinfo {year} {1998})}\BibitemShut {NoStop}%
\end{thebibliography}%
\bibliographystyle{my-asp-style}
\end{document}